\documentclass{agujournal}

\journalname{JGR-Solid Earth}

\begin{document}

%% ------------------------------------------------------------------------ %%
%  Title
%
% (A title should be specific, informative, and brief. Use
% abbreviations only if they are defined in the abstract. Titles that
% start with general keywords then specific terms are optimized in
% searches)
%
%% ------------------------------------------------------------------------ %%

% Example: \title{This is a test title}

\title{Seismic Wave Equations in Tight Oil/Gas Sandstone Media}

%% ------------------------------------------------------------------------ %%
%
%  AUTHORS AND AFFILIATIONS
%
%% ------------------------------------------------------------------------ %%

% Authors are individuals who have significantly contributed to the
% research and preparation of the article. Group authors are allowed, if
% each author in the group is separately identified in an appendix.)

% List authors by first name or initial followed by last name and
% separated by commas. Use \affil{} to number affiliations, and
% \thanks{} for author notes.
% Additional author notes should be indicated with \thanks{} (for
% example, for current addresses).

% Example: \authors{A. B. Author\affil{1}\thanks{Current address, Antartica}, B. C. Author\affil{2,3}, and D. E.
% Author\affil{3,4}\thanks{Also funded by Monsanto.}}

\authors{Jinghuai Gao\affil{1,4}, Weimin Han\affil{2,3}, Haixia Zhao\affil{3,4}, Hui Li\affil{1,4}, Yijie Zhang\affil{1,4}, Jigen Peng\affil{3}, Zongben Xu\affil{3}}

% \affiliation{1}{First Affiliation}
% \affiliation{2}{Second Affiliation}
% \affiliation{3}{Third Affiliation}
% \affiliation{4}{Fourth Affiliation}

\affiliation{1}{School of Electronic and Information Engineering, Xi'an Jiaotong University, Xi'an 710049, China}
\affiliation{2}{Department of Mathematics, University of Iowa, Iowa City, IA 52242, USA}
\affiliation{3}{School of Mathematics and Statistics, Xi'an Jiaotong University, Xi'an 710049, China}
\affiliation{4}{National Engineering Laboratory for Offshore Oil Exploration, Xi'an Jiaotong University, Xi'an 710049, China}

%(repeat as many times as is necessary)

%% Corresponding Author:
% Corresponding author mailing address and e-mail address:

% (include name and email addresses of the corresponding author.  More
% than one corresponding author is allowed in this LaTeX file and for
% publication; but only one corresponding author is allowed in our
% editorial system.)

% Example: \correspondingauthor{First and Last Name}{email@address.edu}

\correspondingauthor{Jinghuai Gao, Haixia Zhao}{jhgao@mail.xjtu.edu.cn; haixia\_zhao@mail.xjtu.edu.cn}

%% Keypoints, final entry on title page.

% Example:
% \begin{keypoints}
% \item	List up to three key points (at least one is required)
% \item	Key Points summarize the main points and conclusions of the article
% \item	Each must be 100 characters or less with no special characters or punctuation
% \end{keypoints}

%  List up to three key points (at least one is required)
%  Key Points summarize the main points and conclusions of the article
%  Each must be 100 characters or less with no special characters or punctuation

\begin{keypoints}
\item A combined system of macro-scale equations is proposed to characterize wave propagation in tight oil/gas sandstone
\item The proposed equations are simpler than Biot equations and more suitable for inversion of porous medium parameters
\item The Goloshubin's diffusive-viscous wave equations can be derived from the proposed equations
\end{keypoints}

%% ------------------------------------------------------------------------ %%
%
%  ABSTRACT
%
% A good abstract will begin with a short description of the problem
% being addressed, briefly describe the new data or analyses, then
% briefly states the main conclusion(s) and how they are supported and
% uncertainties.
%% ------------------------------------------------------------------------ %%

%% \begin{abstract} starts the second page

\begin{abstract}
The paper is devoted to the derivation of a combined system of motion equations for
solid and fluid in isotropic tight oil/gas sandstone media through volume averaging
theorems (VAT). Based on the features of the media, four physical assumptions are proposed
as the foundation for our derivation. More precisely, volume averaging theorems are
applied to the micro-scale motion equations for both the solid and the fluid as well as
to the stress-strain relations, resulting in a combined system of macro-scale equations
for the tight oil/gas sandstone media. It is worth noting that the four 
assumptions may not be satisfied in the whole region. Nevertheless, since the characteristic
diameter for applying VAT ranges between $10^{-6}$ meters and dozens of meters, we may split the entire
domain into several sub-domains such that the four physical assumptions are satisfied in each sub-domain. 
By choosing a proper characteristic diameter of an averaging volume, we derive a
formula for the fluid average pressure in terms of the divergence of the average
displacement from the continuity equation of the fluid. As a result, the motion
equations derived in this paper are simpler than the Biot equations, and are more
suitable for inversion of porous medium parameters. When the fluid is gas and the
compressional wave is considered, the derived motion equations can be simplified to
the diffusive-viscous wave equation. Moreover, the explicit relationship between the
coefficients in this equation and medium parameters is very important for gas
detection in tight gas sandstone.
\end{abstract}

%% ------------------------------------------------------------------------ %%
%
%  TEXT
%
%% ------------------------------------------------------------------------ %%

%%% Suggested section heads:
% \section{Introduction}
%
% The main text should start with an introduction. Except for short
% manuscripts (such as comments and replies), the text should be divided
% into sections, each with its own heading.

% Headings should be sentence fragments and do not begin with a
% lowercase letter or number. Examples of good headings are:

% \section{Materials and Methods}
% Here is text on Materials and Methods.
%
% \subsection{A descriptive heading about methods}
% More about Methods.
%
% \section{Data} (Or section title might be a descriptive heading about data)
%
% \section{Results} (Or section title might be a descriptive heading about the
% results)
%
% \section{Conclusions}

\section{Introduction}

%Text here ===>>>
Wave propagation theory in multi-phase media is the theoretical foundation for oil and gas exploration.
In order to establish the wave propagation theory in multi-phase media \citep{PGM92}, two approaches 
have be used:  one is that applying the concept and theory of macroscopic continuum mechanics into
observable macroscopic quantity; the other is that building macroscopic equations of motion using 
volume averaging technique from microscopic equations satisfied by the fluid and solid grains.

Biot's theory (Biot's model) is based on the first approach \citep{Biot56I, Biot56II}, which is 
one of the most fundamental wave propagation theories in porous media and has been applied widely
in the seismic exploration. 
Biot's theory describes the wave propagation in fluid-saturated porous media. \citet{Biot56I, Biot56II} ignored the microscopic level, assumed that continuum mechanics 
can be used in measurable macroscopic quantity, and derived the governing equations of wave propagation 
using Lagrangian equations. Biot's theory predicted three waves exist in porous media, that is, 
fast P wave, slow P wave and S wave. That prediction was confirmed by experiments, which demonstrates 
the usefulness of Biot's theory \citep{Plo80}. In 1962, Biot investigated the acoustic wave propagation in
porous media and extended the theory to cover heterogenous, anisotropic and viscoelastic media.
Thus, the prototype of wave propagation theory in viscoelastic porous media was established 
\citep{Biot62}.  Based on the theory of irreversible thermodynamics and viscoelasticity, Biot developed the
nonlinear and semilinear mechanics of porous solids \citep{Biot73}.

It was found that Biot's theory could not explain the dispersion and attenuation of waves in ultrasonic 
frequency band. In order to handle this issue, a new model known as Biot and squirt flow (BISQ) 
model was promoted \citep{JTT79, SNWK97,Win85}. It was also found that Biot's theory 
could not be used to describe the attenuation in seismic frequency band, and thus,
the patchy model was proposed \citep{MMD2009}.

Parra extended the BISQ theory for non-isotropic media, and analysed the dispersion of velocity and wave
attenuation in the media \citep{Landau1986,Whit1999}.  Diallo and Appel modified the BISQ theory, 
in which the fluid pressure is independent with squirt flow length \citep{DA00}. Diallo and Prasad
used pulse transmission technique to measure the velocity and attenuation of ultrasonic P and S waves
from two sets of rock samples. Compared with the experimental results on the velocity and attenuation,
it was found that the modified BISQ model provided better prediction than the BISQ model \citep{DPA03}.  
Cheng et al.\ extended the BISQ model for viscoelastic media \citep{TCX12,Yuan-Feng02}.

We now briefly review the literature on wave propagation theories in porous media based on 
the second approach. Cruz and Spanos \citep{DS85} proposed a complete system of equations 
which describes the seismic wave propagation at a low frequency 
in porous media filled with fluid. The method was based on volume-averaging, which combines the 
order-of-magnitude analysis and physical arguments. They obtained equations similar in form to Biot's equations,
and were able to establish relations between the parameters in 
Biot's theory and physical parameters. Considering the interaction between temperature and 
mechanical motion, Cruz and Spanos \citep{DS89} used the volume-averaging technique and proposed governing 
equations in fluid-filled porous media. Pride et al.\ derived linear dynamic equations and stress-strain 
relation in isotropic two-phase (solid and fluid) media using the volume-averaging technique \citep{PGM92}. The macroscopic 
equations matched the equations of motion and stress-strain relation in Biot's theory. 
The effective fluid density is clearly defined with the tractive effort on the interface between pores and 
fluid (the wall of pores) \citep{PGM92}. Cruz et al.\ investigated the  equilibrium thermodynamics in porous 
media based on the volume-averaging technique \citep{CSS93}. Sahay et al.\ considered porous media 
composed of interconnected pores and chemically inert viscous fluid \citep{SSC00}. Averaging the constitutive equation 
at the pore scale, macroscopic constitutive equation in heterogeneous anisotropic porous media was 
developed, and the relationship between this equation and \citep{Biot62} was established \citep{SSC00}. 
Sahay investigated the seismic wave propagation in heterogeneous isotropic media, and developed 
the macroscopic equations of motion and constitutive equation in heterogeneous isotropic media. In 
the most general case, there are twenty-seven independent parameters in the equations \cite{Sah01}. 
Spanos et al.\ derived the complete Biot's theory, taking into consideration of diffusion 
and inertial terms through the volume-averaging theory \citep{SUD02}. Compared with the classical Biot-Gassmann equation, 
the improvement from the complete Biot's theory is that the porosity is a state variable related to
the temperature and time \citep{DS85, HSD95, SUD02}. Spanos investigated seismic
wave propagation in the combinatorial elastic media using volume-averaging theory \citep{S09}.

In summary, much work has been done in the investigation of the porous media and various equations
were proposed. These equations are complicated and contain numerous parameters. The inverse problems
of these equations are improperly posed. For the particular case of oil and gas reservoirs, the
principal factors should be considered and the secondary factors ignored in order to invert
the physical parameters of the underground. 
Using the simplified equations to invert the parameters is a feasible approach \citep{BCZ87}.

Tight oil and gas is an important type of oil and gas, and is widely distributed all over the world in 
different forms such as tight sandstone oil and gas reservoirs, shale oil and gas reservoirs and so on. 
Compared with conventional two-phase media, the matrix and saturated fluid of 
tight oil and gas have special features. In this paper, corresponding to properties of tight oil and gas, 
we propose four basic physical assumptions for tight oil and gas. Based on these assumptions, 
the model is suitably divided into several sub-domains. In each sub-domain, a proper volume-averaging 
characteristic length can be selected. For the microscopic equations of motion (effective equations at each 
point) in each phase (solid and fluid) and stress-strain relation, the macroscopic equations (averaging 
equation) can be derived when the Slattery's volume-averaging technique is used. The detailed derivation 
is given, and the simplification made in the derivation is carefully analysed. As a result, 
conditions for the application of the equations are explicitly stated, and the physical foundation 
of the equations is relatively solid. The equations of motion  in tight oil and gas media are 
two coupled equations, one for solid grains whereas the other for the fluid. For locally 
homogeneous porous media, we present equations for the P and S waves. The governing equations 
in tight oil and gas media derived in this paper are compared with the commonly seen governing equations 
for porous media. It is worth pointing out that the equations for P wave can be transformed 
into the diffusive-viscous wave equations when the fluid in the pores consists of gas only. In addition,
closed form relations are established between the parameters in the equations and the physical parameters 
in tight reservoirs. In existing literature, the diffusive-viscous wave equations were proposed based on 
experiments, and relationship between the coefficients and physical parameters of the reservoirs is not clear.

This paper is organized as follows. In section~\ref{sec:2}, we review the related theory and mathematical tools, 
including the equations of motion of the fluid at pore scale and elastic solid grains in tight oil and gas
reservoirs, and the volume-averaging technique. In section~\ref{sec:3}, we derive the macroscopic equations 
of motion of fluid and solid in tight oil and gas reservoirs, then the equations of motion in tight oil and gas 
reservoirs are achieved. In section~\ref{sec:4}, from the equations of motion in tight oil and gas reservoirs,
we derive the governing equations of P and S waves in locally homogeneous media. In particular, when the 
fluid in pores is gas, the diffusive-viscous wave equations can be obtained. In addition, closed form relations
between the parameters in the equations and the physical parameters in tight oil and gas reservoirs
are given. In section~\ref{sec:5}, we compare the equations proposed in this paper with several commonly used 
wave equations. Discussion and concluding remarks are presented in section~\ref{sec:6}.

\section{Microscopic Motion Equations in Fluid and in Solid}\label{sec:2}

\subsection{Generalized Newton's law of viscous fluids}

Stokes postulated the following fundamental properties of fluids \citep{Ser59}:

(i) The stress tensor is a continuous function of the deformation tensor and the
local thermodynamic state, but independent of other kinematic quantities.

(ii) The fluid is homogeneous and isotropic.

(iii) When there is no deformation, the stress is hydrostatic.

In particular, when the stress tensor is a linear function of the deformation tensor, the 
fluid is called Newtonian.  Experimental results indicate
that the above listed assumptions are reasonable for most fluids and gas \citep{Landau1987}. 
In this paper, we consider porous fluids in the tight oil/gas sandstone that satisfy these assumptions.
The stress-strain relation satisfies the following relation
\begin{linenomath*}
\begin{equation}
\sigma_{ij}^f=\mu_f\left(v^f_{i,j}+v^f_{j,i}-\frac{2}{3}\,v^f_{l,l}\delta_{ij}\right),
\label{2.1}
\end{equation}
\end{linenomath*}
where $\sigma^f=(\sigma_{ij}^f)$ is the viscous stress tensor of the fluid, $\mu_f$ is the dynamic viscosity 
coefficient, and $v^f_i$ is the $i^{th}$ component of fluid velocity ${\mbox{\boldmath{$v$}}}^f$.   Throughout the paper,
we adopt the summation convention over a repeated index. Thus, e.g., 
\[ v^f_{l,l}=\sum_{l=1}^3 \frac{\partial v^f_l}{\partial x_l}. \]

\subsection{Motion equation of fluid in tight oil/gas sandstone}
The flux density of momentum of the viscous fluid can be expressed as \citep{Landau1987}
\begin{linenomath*}
\begin{equation}
 \Pi_{ij}=p^f\delta_{ij}+\rho^f v^f_i v^f_j-\sigma^f_{ij}
\label{2.2}
\end{equation}
\end{linenomath*}
where $p^f$ is the flow pressure and $\rho^f$ is the density of fluid.

According to the Euler equation
\begin{linenomath*}
\begin{equation}
\frac{\partial}{\partial t} \rho^f v^f_i=-\frac{\partial \Pi_{ij}}{\partial x_j}.
\label{2.3}
\end{equation}
\end{linenomath*}
Then the motion equation of fluid can be obtained by combining the equations (\ref{2.2}) and (\ref{2.3}):
\begin{linenomath*}
\begin{equation}
\frac{\partial}{\partial t} \rho^f v^f_i
+\frac{\partial}{\partial x_j}\left(p^f\delta_{ij}+\rho^f v^f_i v^f_j-\sigma^f_{ij}\right)=0,
\label{2.4}
\end{equation}
\end{linenomath*}
where $\sigma_{ij}^f$ is given by equation (\ref{2.1}).

\subsection{Motion equation of solid}
We use ${\mbox{\boldmath{$u$}}}^s  = \left( {u_i^s } \right)$ for the displacement of the solid.
If the thermal effect can be ignored in the mechanical process, and if the body force is negligible,
then the motion equation of solid particle can be expressed as
\begin{linenomath*}
\begin{equation}
\rho^s\frac{\partial^2}{\partial t^2}u^s_i=\frac{\partial \sigma_{ij}^s}{\partial x_j},
\label{2.5}
\end{equation}
\end{linenomath*}
where $\sigma^s=(\sigma_{ij}^s)$ is the stress tensor of the solid:
\begin{linenomath*}
\begin{equation}
\sigma _{ij}^s  = K^s u_{ll}^s \delta _{ij}  + 2\,\mu _s \left( {\varepsilon _{ij} \left( {{\mbox{\boldmath{$u$}}}^s } \right) - \frac{1}{3}u_{ll}^s \delta _{ij} } \right)
\label{2.6}
\end{equation}
\end{linenomath*}
where $K^s$ is the bulk modulus of solid, $\mu_s$ is the shear modulus of solid, ${\mbox{\boldmath{$\varepsilon$}}}\left( {{\mbox{\boldmath{$u$}}}^s } \right) = \left( {\varepsilon _{ij} \left( {{\mbox{\boldmath{$u$}}}^s } \right)} \right)$ is the strain tensor with the components defined by
\begin{linenomath*}
\begin{equation}
\varepsilon _{ij} \left( {{\mbox{\boldmath{$u$}}}^s } \right) = \frac{1}{2}\left( {u_{i,j}^s  + u_{j,i}^s } \right)
\label{2.7}
\end{equation}
\end{linenomath*}
and $u_{ll}^s : = {\rm{tr}}\left( {{\mbox{\boldmath{$\varepsilon$}}}\left( {{\mbox{\boldmath{$u$}}}^s } \right)} \right) = \nabla \cdot{\mbox{\boldmath{$u$}}}^s$ is the divergence of ${\mbox{\boldmath{$u$}}}^s$.

\subsection{The fluid-solid interface condition}

Assume that there is no slip on the interface of the fluid and the solid, 
\begin{linenomath*}
\begin{equation}
{\mbox{\boldmath{$v$}}}^f  = \frac{{\partial {\mbox{\boldmath{$u$}}}^s }}{{\partial t}}.
\label{2.8}
\end{equation}
\end{linenomath*}
Also assume that the normal stress of the solid and the fluid is continuous across the interface, 
\begin{linenomath*}
\begin{equation}
-p^fn_i+\sigma^f_{ij}n_j=\sigma^s_{ij}n_j,
\label{2.9}
\end{equation}
\end{linenomath*}
where ${\mbox{\boldmath{$n$}}} = \left( {n_i } \right)$ is the unit normal vector on the interface.

\subsection{Averaging formulas}

\subsubsection{Volume averaging and phase averaging}

We will take volume averaging over particular regions in porous media, which are of the same shape, 
volume and orientation. In this paper, we will use balls of a constant radius $r$ as the particular 
regions to do the averaging. For simplicity, we use $V\left( {\mbox{\boldmath{$x$}}} \right): = B\left( {{\mbox{\boldmath{$x$}}};r} \right)$ to denote the ball centered 
at ${\mbox{\boldmath{$x$}}}$ with radius $r$, and use $|V|$ for the volume of $V\left( {\mbox{\boldmath{$x$}}} \right)$.  Note that $|V|$ does not 
depend on ${\mbox{\boldmath{$x$}}}$.

Denote by $G^f$ for a physical quantity related to the fluid, and extend the definition of $G^f$ 
to take on the value $0$ outside the fluid region.  We consider two kinds of averaging: volume averaging 
and phase averaging. The volume average of $G^f$ over $V\left( {\mbox{\boldmath{$x$}}} \right)$ is defined as
\begin{linenomath*}
\begin{equation}
\langle G^f \rangle \left( \mbox{\boldmath{$x$}} \right): = \frac{1}{{\left| V \right|}}\int_{V\left( {\mbox{\boldmath{$x$}}} \right)} {G^f \left( {{\mbox{\boldmath{$x'$}}}} \right)d{\mbox{\boldmath{$x'$}}}}.
\label{2.10}
\end{equation}
\end{linenomath*}
If the grain size of the solid is greater than $r$, we may assume $\langle G^f \rangle \left( x \right)$ is a continuous function. 

Let $V^f \left( {\mbox{\boldmath{$x$}}} \right) \subset V\left( {\mbox{\boldmath{$x$}}} \right)$ be the subset of $V\left( {\mbox{\boldmath{$x$}}} \right)$ occupied by the fluid.  
The phase average of $G^f$ over $V\left( {\mbox{\boldmath{$x$}}} \right)$ is then defined by
\begin{linenomath*}
\begin{equation}
\overline {G^f } \left( {\mbox{\boldmath{$x$}}} \right): = \frac{1}{{\left| {V^f \left( {\mbox{\boldmath{$x$}}} \right)} \right|}}\int_{V\left( {\mbox{\boldmath{$x$}}} \right)} {G^f \left( {{\mbox{\boldmath{$x'$}}}} \right)d{\mbox{\boldmath{$x'$}}}}.
\label{2.11}
\end{equation}
\end{linenomath*}

We define the porosity  of the region $V\left( {\mbox{\boldmath{$x$}}} \right)$ by the formula
\begin{linenomath*}
\begin{equation}
\phi \left( {\mbox{\boldmath{$x$}}} \right): = \frac{{\left| {V^f \left( {\mbox{\boldmath{$x$}}} \right)} \right|}}{{\left| V \right|}}.
\label{2.12}
\end{equation}
\end{linenomath*}
In general, $\phi \left( {\mbox{\boldmath{$x$}}} \right)$ is a function of the location $\mbox{\boldmath{$x$}}$ (and the radius $r$).  For simplicity,
we will assume $\phi \left( {\mbox{\boldmath{$x$}}} \right) = \phi$ to be independent of the location $\mbox{\boldmath{$x$}}$  in the rest of the paper. 
From the definition of the porosity, we observe that
\begin{linenomath*}
\begin{equation}
\left\langle {G^f } \right\rangle \left( {\mbox{\boldmath{$x$}}} \right) = \phi \overline {G^f } \left( {\mbox{\boldmath{$x$}}} \right).
\label{2.13}
\end{equation}
\end{linenomath*}

Similarly, for a physical quantity related to the solid, we define its value to be $0$ in the fluid 
region, and define its volume average over $V\left( {\mbox{\boldmath{$x$}}} \right)$ to be
\[\langle G^s \rangle \left( \mbox{\boldmath{$x$}} \right): = \frac{1}{{\left| V \right|}}\int_{V\left( {\mbox{\boldmath{$x$}}} \right)} {G^s \left( {{\mbox{\boldmath{$x'$}}}} \right)d{\mbox{\boldmath{$x'$}}}} .\]
The phase average of $G^s$ over $V(\mbox{\boldmath{$x$}})$ is 
\[\overline {G^s } \left( {\mbox{\boldmath{$x$}}} \right): = \frac{1}{{\left| {V^s \left( {\mbox{\boldmath{$x$}}} \right)} \right|}}\int_{V\left( {\mbox{\boldmath{$x$}}} \right)} {G^s \left( {{\mbox{\boldmath{$x'$}}}} \right)d{\mbox{\boldmath{$x'$}}}} \]
where $V^s \left( {\mbox{\boldmath{$x$}}} \right) \subset V\left( {\mbox{\boldmath{$x$}}} \right)$ is the subset of $V\left( {\mbox{\boldmath{$x$}}} \right)$ occupied by the solid. Note that 
\[ V\left( {\mbox{\boldmath{$x$}}} \right) = V^f \left( {\mbox{\boldmath{$x$}}} \right) \cup V^s \left( {\mbox{\boldmath{$x$}}} \right),\quad \left| {V^f \left( {\mbox{\boldmath{$x$}}} \right)} \right| + \left| {V^s \left( {\mbox{\boldmath{$x$}}} \right)} \right| = \left| V \right|.\]

We will make use of some volume averaging formulas. To state the formulas, we denote by 
$A^{fs} \left( {\mbox{\boldmath{$x$}}} \right)$ the interface of the solid and fluid in $V\left( {\mbox{\boldmath{$x$}}} \right)$, by ${\mbox{\boldmath{$v$}}}$ the fluid velocity on $A^{fs} \left( {\mbox{\boldmath{$x$}}} \right)$, by ${\mbox{\boldmath{$n$}}}$ the unit outward normal vector of the fluid on $A^{fs} \left( {\mbox{\boldmath{$x$}}} \right)$.

The first formula is a relation between the time derivative of the average and the average
of the time derivative \citep{FZ2005, PGM92}: let $G^f \left( {{\mbox{\boldmath{$x$}}},t} \right)$ be a function of ${\mbox{\boldmath{$x$}}}$ and $t$,
\begin{linenomath*}
\begin{equation}
\partial _t \left\langle {G^f } \right\rangle \left( {{\mbox{\boldmath{$x$}}},t} \right) = \left\langle {\partial _t G^f } \right\rangle \left( {{\mbox{\boldmath{$x$}}},t} \right) + \frac{1}{{\left| V \right|}}\int_{A^{fs} \left( {\mbox{\boldmath{$x$}}} \right)} {G^f \left( {{\mbox{\boldmath{$x'$}}},t} \right)v\left( {{\mbox{\boldmath{$x'$}}},t} \right) \cdot {\mbox{\boldmath{$n$}}}\left( {{\mbox{\boldmath{$x'$}}}} \right)ds\left( {{\mbox{\boldmath{$x'$}}}} \right)}.
\label{2.14a}
\end{equation}
\end{linenomath*}

The second formula is a relation between the spatial derivative of the average and the average
of the spatial derivative \citep{FZ2005, Whit1999}:
\begin{linenomath*}
\begin{equation}
\left\langle {\partial _i G^f } \right\rangle \left( {{\mbox{\boldmath{$x$}}},t} \right) = \partial _i \left\langle {G^f } \right\rangle \left( {{\mbox{\boldmath{$x$}}},t} \right) + \frac{1}{{\left| V \right|}}\int_{A^{fs} \left( {\mbox{\boldmath{$x$}}} \right)} {G^f \left( {{\mbox{\boldmath{$x'$}}},t} \right)n_i \left( {{\mbox{\boldmath{$x'$}}}} \right)ds\left( {{\mbox{\boldmath{$x'$}}}} \right)} .
\label{2.15a}
\end{equation}
\end{linenomath*}

Taking $G^f \left( {\mbox{\boldmath{$x$}}} \right) = 1$ in the formula (\ref{2.15a}), we obtain an important relation:
\begin{linenomath*}
\begin{equation}
\partial _i \phi \left( {\mbox{\boldmath{$x$}}} \right) =  - \frac{1}{{\left| V \right|}}\int_{A^{fs} \left( {\mbox{\boldmath{$x$}}} \right)} {n_i \left( {{\mbox{\boldmath{$x'$}}}} \right)ds\left( {{\mbox{\boldmath{$x'$}}}} \right)} .
\label{2.16a}
\end{equation}
\end{linenomath*}

Another consequence of the formula (\ref{2.15a}) is a relation between the divergence of the average
and the average of the divergence \citep{FZ2005, Whit1999}:
\begin{linenomath*}
\begin{equation}
\left\langle {\nabla  \cdot {\mbox{\boldmath{$\psi$}}}} \right\rangle \left( {{\mbox{\boldmath{$x$}}},t} \right) = \nabla  \cdot \left\langle {\mbox{\boldmath{$\psi$}}} \right\rangle \left( {{\mbox{\boldmath{$x$}}},t} \right) + \frac{1}{{\left| V \right|}}\int_{A^{fs} \left( {\mbox{\boldmath{$x$}}} \right)} {{\mbox{\boldmath{$\psi$}}}\left( {{\mbox{\boldmath{$x'$}}},t} \right) \cdot {\mbox{\boldmath{$n$}}}\left( {{\mbox{\boldmath{$x'$}}}} \right)ds\left( {{\mbox{\boldmath{$x'$}}}} \right)} .
\label{2.17a}
\end{equation}
\end{linenomath*}

In the rest of the paper, to simplify the writing, we will follow the convention in the literature and
suppress the independent variables for various quantities involved in the averaging formulas; moreover,
we will use $V$ to replace $|V|$ for the volume of $V$, use $dV$ for the volume element and
$dA$ for the surface element.  Thus, the formulas (\ref{2.14a})--(\ref{2.17a}) are each expressed
as follows:
\begin{linenomath*}
\begin{eqnarray}
\partial _t \left\langle {G^f } \right\rangle  &=& \left\langle {\partial _t G^f } \right\rangle  + \frac{1}{{\left| V \right|}}\int_{A^{fs} } {G^f {\mbox{\boldmath{$v$}}} \cdot {\mbox{\boldmath{$n$}}}dA} ,
\label{2.14}\\
\left\langle {\partial _i G^f } \right\rangle  &=& \partial _i \left\langle {G^f } \right\rangle  + \frac{1}{{\left| V \right|}}\int_{A^{fs} } {G^f n_i dA}  ,
\label{2.15}\\
\partial _i \phi  &=&  - \frac{1}{{\left| V \right|}}\int_{A^{fs} } {n_i dA} ,
\label{2.16}\\
\left\langle {\nabla  \cdot {\mbox{\boldmath{$\psi$}}}} \right\rangle  &=& \nabla  \cdot \left\langle {\mbox{\boldmath{$\psi$}}} \right\rangle  + \frac{1}{{\left| V \right|}}\int_{A^{fs} } {{\mbox{\boldmath{$\psi$}}} \cdot {\mbox{\boldmath{$n$}}}dA} .
\label{2.17}
\end{eqnarray}
\end{linenomath*}

\subsubsection{On proper size choice in volume averaging}

The scale determination is very important in applying the technique of volume averaging \citep{Whit1999}. 
If the size of representative volume is too small to include both 
solid phase and fluid phase, then the average value could be zero. Consequently, the average 
value may have only the zero-th order smoothness, and its gradient is not defined. However,
the idea of the averaging method is to define a function $\langle G\rangle$ smoother than the function 
$G$ \citep{PGM92}. Thus, letting $s$ represent an average length of the grain and letting 
$l=2\,r$ be the diameter of the balls, in order to make 
the averaging results useful, we require the following condition
\begin{linenomath*}
\begin{equation}
s\ll l.
\label{2.18}
\end{equation}
\end{linenomath*}
On the other hand, when a wave passes through the porous media, to get effective value after 
volume averaging, we require
\begin{linenomath*}
\begin{equation}
l\ll \lambda,
\label{2.19}
\end{equation}
\end{linenomath*}
where $\lambda$ denotes the wavelength. Combining (\ref{2.18}) and (\ref{2.19}), we reach the condition
on the averaging volume size:
\begin{linenomath*}
\begin{equation}
s\ll l\ll \lambda,
\label{2.20}
\end{equation}
\end{linenomath*}

In tight oil/gas sandstone media, $s$ is of the order $10^{-6}$\,m (cf.\ Appendix B), the compressional 
wave speed is in the range of 3800\,m/s to 5570\,m/s, whereas the shear wave speed 
is in the range of 2590\,m/s to 3500\,m/s.  Assume the dominant frequency of the 
seismic wave is 40\,Hz.  Then the wavelength of the compressional wave, $\lambda_P$, is between
95\,m and 139.25\,m, whereas that of the shear wave, $\lambda_S$, is between
64.75\,m and 87.5\,m.  Thus, in meters, for the compressional wave, 
\[ 10^{-6}\ll l\ll 95,\]
and for the shear wave, 
\[ 10^{-6}\ll l\ll 64.75.\]
Hence, the range of the characteristic length (the diameter of the balls in this paper) is wide 
for the application of the volume averaging formulas.  Based on the need from applications, 
it is also possible to choose averaging volumes of different characteristic sizes in different
sub-regions of the media.

\section{Motion Equations for Tight Oil/Gas Sandstone Media}\label{sec:3}

\subsection{Basic physical assumptions on oil/gas sandstone media}\label{subsec:31}
Physical assumptions are needed in deriving the wave motion equation by using the averaging
formulas \citep{PGM92}. In this paper, we introduce the following four basic assumptions on the tight
oil/gas sandstone media.

\noindent\underline{Assumption 1}.  Both the fluid and solid may be viewed homogeneous in a scale much 
larger than the characteristic pore size, but much smaller than the wavelength of 
seismic wave.  Thus, for either the fluid phase or the solid phase, over the sub-region $V$ 
where averaging formulas are applied, material parameters such as the density may be taken 
as constants.

\noindent\underline{Assumption 2}.  Both the fluid and solid are isotropic, and the thermodynamic
effect can be ignored.

\noindent\underline{Assumption 3}.  Both the fluid and solid are in equilibrium before the seismic wave arrives.

\noindent\underline{Assumption 4}.  The porosity gradient in the tight oil/gas sandstone is approximately zero 
on a scale several times bigger than the size of $V$ before the seismic wave arrives.

For the oil/gas sandstone media under consideration, wherever necessary, we can split the entire domain
into sub-domains so that on each sub-domain, the above four basic assumptions are valid.  We then derive
motion equations in each sub-domain, and combine them together with joining conditions across the 
boundaries of the sub-domains, so as to form a system of equations and conditions over the entire domain.
For simplicity in writing, in Subsections~\ref{subsec:32}--\ref{subsec:34}, we suppress the sub-domain index 
when referring to various quantities.

\subsection{Macro-scale equations for the fluid}\label{subsec:32}

First, we recall the poro-scale fluid flow equations (\ref{2.4}) and (\ref{2.1}),
\begin{linenomath*}
\begin{eqnarray}
&\partial_t\left(\rho^fv^f_i\right)+\partial_j\left(p^f\delta_{ij}+\rho^f v^f_i v^f_j-\sigma^f_{ij}\right)=0,\label{3.1}\\
&\sigma_{ij}^f=\mu_f\left(v^f_{i,j}+v^f_{j,i}-\frac{2}{3}\,v^f_{l,l}\delta_{ij}\right).
\label{3.2}
\end{eqnarray}
\end{linenomath*}
Taking the volume average on the equation (\ref{3.1}) over $V = V({\mbox{\boldmath{$x$}}})$, we obtain
\begin{linenomath*}
\begin{equation}
\left\langle\partial_t\left(\rho^f v^f_i\right)\right\rangle+
\left\langle\partial_j\left(p^f\delta_{ij}+\rho^f v^f_i v^f_j-\sigma^f_{ij}\right)\right\rangle=0.
\label{3.3}
\end{equation}
\end{linenomath*}
By the formula (\ref{2.14}),
\begin{linenomath*}
\begin{equation}
\left\langle\partial_t\left(\rho^f v^f_i\right)\right\rangle = \partial_t\left\langle\rho^f v^f_i\right\rangle
-\frac{1}{V}\int_{A^{fs}}\rho^f v^f_i v^f_j n_j dA.
\label{3.4}
\end{equation}
\end{linenomath*}
By the formula (\ref{2.15}),
\begin{linenomath*}
\begin{eqnarray}
\left\langle\partial_j\left(p^f\delta_{ij}+\rho^f v^f_i v^f_j-\sigma^f_{ij}\right)\right\rangle &=&\partial_j\left(\left\langle p^f\delta_{ij}\right\rangle+\left\langle\rho^f v^f_i v^f_j\right\rangle -\left\langle\sigma^f_{ij}\right\rangle\right)\nonumber\\
&&{}\quad +\frac{1}{V}\int_{A^{fs}} \left(p^f\delta_{ij}+\rho^f v^f_i v^f_j-\sigma^f_{ij}\right)n_j dA.
\label{3.5}
\end{eqnarray}
\end{linenomath*}
Use (\ref{3.4}) and (\ref{3.5}) in (\ref{3.3}) to obtain
\begin{linenomath*}
\begin{equation}
\partial_t\left\langle\rho^f v^f_i\right\rangle+\partial_j\left(\left\langle p^f\delta_{ij}\right\rangle
+\left\langle\rho^f v^f_i v^f_j\right\rangle-\left\langle\sigma^f_{ij}\right\rangle\right)
+\frac{1}{V}\int_{A^{fs}} \left(p^f\delta_{ij}-\sigma^f_{ij}\right)n_j dA=0.
\label{3.6}
\end{equation}
\end{linenomath*}
Making use of the definition of porosity (\ref{2.11}), we rewrite the equation (\ref{3.6}) as
\begin{linenomath*}
\begin{equation}
	\partial_t\left(\phi\,\overline{\rho^f v^f_i}\right)+\partial_j\left[\phi\,\overline{p^f}\delta_{ij}
	+\phi\,\overline{\rho^f v^f_i v^f_j}-\left\langle\sigma^f_{ij}\right\rangle\right]
	+\frac{1}{V}\int_{A^{fs}} \left(p^f\delta_{ij}-\sigma^f_{ij}\right)n_j dA=0.
	\label{3.7}
\end{equation}
\end{linenomath*}

Taking the volume average on the equation (\ref{3.2}) over $V=V({\mbox{\boldmath{$x$}}})$, we obtain
\begin{linenomath*}
\begin{equation}
	\left\langle\sigma_{ij}^f\right\rangle=\mu_f\left[\left\langle v^f_{i,j}\right\langle+\left\langle v^f_{j,i}\right\rangle
	-\frac{2}{3}\left\langle v^f_{l,l}\right\rangle\delta_{ij}\right].
	\label{3.8}
\end{equation}
\end{linenomath*}
Then apply the averaging formulas to find
\begin{linenomath*}
\begin{equation}
	\left\langle\sigma_{ij}^f\right\rangle=\mu_f\left[\partial_j\left\langle v^f_{i}\right\rangle
	+\partial_i\left\langle v^f_{j}\right\rangle-\frac{2}{3}\,\partial_l\left\langle v^f_{l}\right\rangle\delta_{ij}
	+\frac{1}{V}\int_{A^{fs}} \left(v^f_j n_i+v^f_i n_j-\frac{2}{3}\, v^f_l n_l\delta_{ij}\right) dA \right].
	\label{3.9}
\end{equation}
\end{linenomath*}

According to \citep{DS85}, up to a higher order term, 
\begin{linenomath*}
\begin{equation}
	\frac{1}{V}\int_{A^{fs}} \left(v^f_j n_i+v^f_i n_j-\frac{2}{3}\, v^f_l n_l\delta_{ij}\right) dA
	=-\overline{v^f_j}\partial_i\phi-\overline{v^f_i}\partial_j\phi
	+\frac{2}{3}\,\overline{v^f_l}\partial_l\phi \delta_{ij}.
	\label{3.10}
\end{equation}
\end{linenomath*}
Since by Assumption 4, the gradient of the porosity is approximately zero, the right side of 
(\ref{3.10}) is nearly zero.  Then we derive from (\ref{3.9}) that, up to a higher order term, 
\begin{linenomath*}
\begin{equation}
	\left\langle\sigma_{ij}^f\right\rangle=\mu_f\left[\partial_j\left\langle v^f_{i}\right\rangle
	+\partial_i\left\langle v^f_{j}\right\rangle-\frac{2}{3}\,\partial_l\left\langle v^f_{l}\right\rangle\delta_{ij}\right].
	\label{3.11}
\end{equation}
\end{linenomath*}
Again, make use of the definition of porosity (\ref{2.11}) to rewrite the equation (\ref{3.11}) as
\begin{linenomath*}
\begin{equation}
	\left\langle\sigma_{ij}^f\right\rangle=\mu_f\phi \left(\partial_j\overline{v^f_{i}}
	+\partial_i\overline{v^f_{j}}-\frac{2}{3}\,\partial_l\overline{v^f_{l}}\delta_{ij}\right).
	\label{3.12}
\end{equation}
\end{linenomath*}

Use (\ref{3.12}) in (\ref{3.7}) and ignore the second order term of the velocity components,
\begin{linenomath*}
\begin{equation}
	\partial_t\left(\phi\,\overline{\rho^f v^f_i}\right)+\partial_j\left[\phi\,\overline{p^f}\delta_{ij}
	-\mu_f\phi \left(\partial_j\overline{v^f_{i}}
	+\partial_i\overline{v^f_{j}}-\frac{2}{3}\,\partial_l\overline{v^f_{l}}\delta_{ij}\right)\right]
	+\frac{1}{V}\int_{A^{fs}} \left(p^f\delta_{ij}-\sigma^f_{ij}\right)n_j dA=0.
	\label{3.13}
\end{equation}
\end{linenomath*}
Denote
\[ I_1=\frac{1}{V}\int_{A^{fs}} p^f n_i dA,\qquad I_2=-\frac{1}{V}\int_{A^{fs}} \sigma^f_{ij}n_j dA.\]
Write
\begin{linenomath*}
\begin{equation}
I_1=\frac{1}{V}\int_{A^{fs}} \left(p^f-\overline{p^f}\right) n_i dA
+\frac{1}{V}\int_{A^{fs}} \overline{p^f} n_i dA,
\label{3.14}
\end{equation}
\end{linenomath*}
where $\overline{p^f}$ is constant in $V({\mbox{\boldmath{$x$}}})$.  When the wave induced fluid flow occurs, the pressure 
$p^f$ at the micro-scale deviates from its average value $\overline{p^f}$. For the generalized Newtonian 
fluid flow (i.e., Stokes flow), we may assume \citep{DS83}
\begin{linenomath*}
\begin{equation}
	\frac{1}{V}\int_{A^{fs}} \left(p^f-\overline{p^f}\right) n_i dA
	=\mu_f b\left(\overline{v^f_i}-\overline{v^s_i}\right),
	\label{3.15}
\end{equation}
\end{linenomath*}
where the parameter $b$ is a geometry related parameter in the effective flow tube. Then we obtain 
from (\ref{3.14}) that
\begin{linenomath*}
\begin{equation}
	I_1=\frac{1}{V}\int_{A^{fs}} p^f n_i dA=\mu_f b\left(\overline{v^f_i}-\overline{v^s_i}\right)
	-\overline{p^f}\,\partial_i\phi.
	\label{3.16}
\end{equation}
\end{linenomath*}
Following \citep{DS83},
\begin{linenomath*}
\begin{equation}
	I_2=\mu_f a\left(\overline{v^f_i}-\overline{v^s_i}\right),
	\label{3.17}
\end{equation}
\end{linenomath*}
where $a$ is a geometry related parameter of the effective flow tube.

Using (\ref{3.16}) and (\ref{3.17}), we derive from (\ref{3.13}) that
\begin{linenomath*}
\begin{equation}
	\partial_t\left(\phi\,\overline{\rho^f v^f_i}\right)+\partial_j\left[\phi\,\overline{p^f}\delta_{ij}
	-\mu_f\phi \left(\partial_j\overline{v^f_{i}}
	+\partial_i\overline{v^f_{j}}-\frac{2}{3}\,\partial_l\overline{v^f_{l}}\delta_{ij}\right)\right]
	+\mu_f \left(a+b\right)\left(\overline{v^f_i}-\overline{v^s_i}\right)=0.
	\label{3.18}
\end{equation}
\end{linenomath*}
It can be shown that (cf.\ Appendix A)
\begin{linenomath*}
\begin{equation}
	a+b=\frac{\phi^2}{K},
	\label{3.19}
\end{equation}
\end{linenomath*}
where $K$ is the permeability.  Thus, the equation (\ref{3.18}) can be rewritten as
\begin{linenomath*}
\begin{equation}
	\partial_t\left(\rho^f \overline{v^f_i}\right)+\partial_j\left[\overline{p^f}\delta_{ij}
	-\mu_f\left(\partial_j\overline{v^f_{i}}
	+\partial_i\overline{v^f_j}-\frac{2}{3}\,\partial_l\overline{v^f_l}\delta_{ij}\right)\right]
	+\mu_f \,\frac{\phi}{K}\left(\overline{v^f_i}-\overline{v^s_i}\right)=0.
	\label{3.20}
\end{equation}
\end{linenomath*}
In the vector notation, where $(1/V)\int_{A^{fs}}u^f_i \mbox{\boldmath{$v$}}{\cdot}\mbox{\boldmath{$n$}}\,dA$ is negligible, 
the equation takes the form
\begin{linenomath*}
\begin{equation}
	\rho^f \partial_t^2 \overline{\mbox{\boldmath{$u$}}^f}+\nabla \overline{p^f}
	-\mu_f\left(\frac{1}{3}\,\nabla \nabla{\cdot}\overline{\mbox{\boldmath{$v$}}^f}+\Delta \overline{\mbox{\boldmath{$v$}}^f}\right)
	+\mu_f \,\frac{\phi}{K}\left(\overline{\mbox{\boldmath{$v$}}^f}-\overline{\mbox{\boldmath{$v$}}^s}\right)=0.
	\label{3.21}
\end{equation}
\end{linenomath*}
The equations (\ref{3.20}) or (\ref{3.21}) are the macro-scale fluid equations.

\subsection{Macro-scale equations for the solid}\label{subsec:33}

The motion equation for the solid at pore-scale is described as 
\begin{linenomath*}
\begin{equation}
	\rho^s \partial_t^2 u^s_i=\partial_j \sigma^s_{ij}. 
	\label{3.22}
\end{equation}
\end{linenomath*}
Taking the volume averaging of the equation (\ref{3.22}), we have
\begin{linenomath*}
\begin{equation}
	\langle\rho^s\partial_t^2u_i^s\rangle  = \langle \partial_j\sigma_{ij}^s\rangle
	=\partial_j\langle \sigma_{ij}^s\rangle  + \frac{1}{V}\int_{A^{sf}} \sigma_{ij}^s {n'_j}dA.
	\label{3.23}
\end{equation}
\end{linenomath*}

For the left-side of the equation (\ref{3.23}),
\begin{linenomath*}
\begin{eqnarray}
	\partial _t^2u_i^s & = &\partial_t(\partial_t u_i^s),
	\label{3.24a}\\
	\langle \partial _t^2u_i^s\rangle &=&\langle\partial_t(\partial_t u_i^s)\rangle=\partial_t\langle\partial_t u_i^s\rangle
	- \frac{1}{V}\int_{A^{fs}} \partial_t u_i^s \partial_t u_j^s {n_j}dA.
	\label{3.24b}
\end{eqnarray}
\end{linenomath*}
Ignoring the second-order quantity of the velocity for the solid, we obtain
\begin{linenomath*}
\begin{equation}
	\langle\partial_t^2u_i^s\rangle = \partial_t\langle\partial_t u_i^s\rangle.
	\label{3.25}
\end{equation}
\end{linenomath*}
Apply the volume averaging formula (\ref{3.5}),
\begin{linenomath*}
\begin{equation}
	\langle\partial _t u_i^s\rangle=\partial_t\langle u_i^s\rangle - \frac{1}{V}\int_{A^{fs}} u_i^s \partial_t u_j^s{n_j}dA.
	\label{3.26}
\end{equation}
\end{linenomath*}
The second term on the right side of (\ref{3.26}) is a second-order quantity and is ignored.  
Then we obtain from the equation (\ref{3.25}) that
\begin{linenomath*}
\begin{equation}
	\langle\partial_t^2 u_i^s\rangle=\partial_t^2\langle u_i^s\rangle = (1 -\phi)\partial_t^2 \overline{u_i^s}.
	\label{3.27}
\end{equation}
\end{linenomath*}

We then consider the right-side term in the equation (\ref{3.23}),
\begin{linenomath*}
\begin{equation}
	\left\langle\partial_j\sigma_{ij}^s\right\rangle =\partial_j\left\langle\sigma_{ij}^s \right\rangle  
	+ \frac{1}{V}\int_{A^{sf}} \sigma_{ij}^s{n_j}dA.
	\label{3.28}
\end{equation}
\end{linenomath*}
Substituting the equation (\ref{2.6}) into the expression in equation (\ref{3.28}), we find
\begin{linenomath*}
\begin{equation}
	\left\langle \sigma_{ij}^s \right\rangle  = \left\langle K_s u_{ll}^s\delta_{ij} \right\rangle
	+ 2\mu_s\left\langle u_{ij}^s \right\rangle-\frac{2\mu_s}{3}\left\langle u_{ll}^s\delta_{ij}\right\rangle, 
	\label{3.29}
\end{equation}
\end{linenomath*}
\begin{linenomath*}
\begin{equation}
	\left\langle K_s u_{ll}^s\delta_{ij}\right\rangle =K_s\delta_{ij}\left\langle\partial_l u_l^s\right\rangle 
	=K_s\delta_{ij}\left[\partial_l\left\langle u_l^s \right\rangle +\frac{1}{V}\int_{A^{sf}}u_l^s{n_l}dA \right],
	\label{3.30}
\end{equation}
\end{linenomath*}
\begin{linenomath*}
\begin{equation}
	\left\langle u_{ij}^s\right\rangle\!=\! \frac{1}{2}\left[\left\langle \partial_k u_i^s \right\rangle  
	+ \left\langle\partial_i u_j^s\right\rangle\right] 
	\!=\! \frac{1}{2}\left[\partial_j\left\langle u_i^s\right\rangle+\partial_i\left\langle u_j^s\right\rangle 
	\!+\! \frac{1}{V}\int_{A^{sf}}\!\! u_i^s{n_j}dA  \!+\! \frac{1}{V}\int_{A^{sf}}\!\! u_j^s{n_i}dA\right],
	\label{3.31}
\end{equation}
\end{linenomath*}
\begin{linenomath*}
\begin{equation}
	\left\langle \nabla{\cdot}\mbox{\boldmath{$u$}}^s \right\rangle
	=\nabla\cdot\left\langle \mbox{\boldmath{$u$}}^s \right\rangle + \frac{1}{V}\int_{A^{sf}} u_j^s{n_j}dA
	=\partial_l\left\langle u_l^s \right\rangle  + \frac{1}{V}\int_{A^{sf}} u_j^s {n_j}dA.
	\label{3.32}
\end{equation}
\end{linenomath*}
Substituting equation (\ref{3.30}), (\ref{3.31}) and (\ref{3.32}) into equations (\ref{3.29}), we obtain
\begin{linenomath*}
\begin{eqnarray}
	\left\langle \sigma_{ij}^s\right\rangle &=& K_s\delta_{ij}\left[\partial_l\left\langle u_l^s\right\rangle  
	+ \frac{1}{V}\int_{A^{sf}} u_l^s{n_l}dA\right] - \frac{2}{3}{\mu_s}\delta _{ij}\left(\partial_l\left\langle u_l^s\right\rangle 
	+ \frac{1}{V}\int_{A^{sf}} u_j^s{n_j}dA\right) \nonumber\\
&&+\mu_s\left\{\partial_j\left\langle u_i^s\right\rangle 
	+ \partial_i\left\langle u_j^s \right\rangle+\frac{1}{V}\int_{A^{sf}} \left(u_i^s{n_j}+u_j^s{n_i}\right)dA
	\right\}	.
	\label{3.33}
\end{eqnarray}
\end{linenomath*}
The expression $\mbox{\boldmath{$u$}}^s{\cdot}\mbox{\boldmath{$n$}}\, dA$ represents the volume swept out by the displacement of 
the boundary surface element, and thus it is related to the variation of the porosity. We have
\begin{linenomath*}
\begin{equation}
	\frac{1}{V}\int_{A^{sf}} \mbox{\boldmath{$u$}}^s{\cdot}\mbox{\boldmath{$n$}}\, dA  =  - (\phi  - {\phi_{0}}),
	\label{3.34}
\end{equation} 
\end{linenomath*}
$\phi $ and ${\phi _0}$ being the porosity before and after perturbation, respectively. 
Consider the quantity
\begin{linenomath*}
\begin{equation}
	I=\frac{1}{V}\int_{A^{sf}} {\left( {u_i^s{n_j} + u_j^s{n_i}} \right)dA
		- \frac{2}{3}} {\delta _{ij}}\frac{1}{V}\int_{A^{sf}} {u_l^s{n_l}dA}.
	\label{3.35}
\end{equation} 
\end{linenomath*}
Its time derivative is, noting that $v_i^f = \partial u_i^s/\partial t$ on the interface 
of fluid and solid, 
\begin{linenomath*}
\begin{eqnarray}
	\frac{{\partial I}}{{\partial t}}  
	&=& \frac{1}{V}\int_{A^{sf}} {\left( {v_i^f{n_j} + v_j^f{n_i}} \right)dA
		- \frac{2}{3}} {\delta _{ij}}\frac{1}{V}\int_{A^{sf}} {v_l^f{n_l}dA} \nonumber\\ 
	&=&  - \overline {v_i^f}{\partial _j}\phi  - \overline {v_j^f}{\partial _i}\phi 
	+ \frac{2}{3}{\delta _{ij}}\overline {v_l^f}{\partial _l}\phi 
	\label{3.36}
\end{eqnarray}
\end{linenomath*} 

The particle velocity on the interface between the solid and fluid in tight oil and gas media is relatively small, 
and the porosity varies slowly. Therefore, according to Assumption 4, the right side of (\ref{3.36}) is
nearly zero,
\begin{linenomath*}
\begin{equation}
	\frac{\partial I}{\partial t}\approx 0.
	\label{3.37}
\end{equation}  
\end{linenomath*}
This implies that $I$ is approximately a constant on the interface between the solid and fluid.
Choosing the constant to be zero, we have
\begin{linenomath*}
\begin{equation}
	I \approx 0.
	\label{3.38}
\end{equation}
\end{linenomath*}
Moreover,
\[ \partial _j\langle u_i^s \rangle = \partial_j\frac{1}{V}\int_{V^s}u_i^sdV
= \partial_j\overline {u_i^s}\left(\frac{V^s}{V}\right) 
= \partial_j\overline {u_i^s}\left(\frac{V -V^f}{V}\right)
= \partial_j((1 - \phi )\overline {u_i^s}),\]
i.e.,
\begin{linenomath*}
\begin{equation}
	\partial_i\langle u_j^s\rangle = (1 - \phi ){\partial _i}\overline {u_j^s}.
	\label{3.40}
\end{equation} 
\end{linenomath*}

Using (\ref{3.34}), (\ref{3.38}) and (\ref{3.40}) in (\ref{3.33}), and noting the relation 
${\partial _i}\langle u_j^s \rangle = (1 - \phi ){\partial _i}\overline {u_j^s}$, we have
\begin{linenomath*}
\begin{eqnarray}
	\langle\sigma_{ij}^s\rangle
	&=& {\mu _s}\left\{ (1 - \phi )\overline {u_{i,j}^s} + (1 - \phi )\overline {u_{j,i}^s} 
	- \frac{2}{3}{\delta _{ij}}(1 - \phi )\overline {u_{l,l}^s} \right\} 
	+ K_s\delta_{ij}\left\{ (1 - \phi )\overline {u_{l,l}^s}  + \phi  - {\phi _0}\right\} \nonumber \\
	&=& {\mu _s}(1-\phi )\left\{\partial_j\overline{u_i^s} + \partial_i\overline{u_j^s}
	- \frac{2}{3}\delta_{ij}\nabla{\cdot}\overline{\mbox{\boldmath{$u$}}^s}\right\}  
	+ K_s\delta_{ij}(1 - \phi )\nabla{\cdot}\overline{\mbox{\boldmath{$u$}}^s}  \label{3.41}\\
	&& +\left( K_s- \frac{2}{3}\mu_s \right)\delta_{ij}\left(\phi_{0}-\phi\right). \nonumber	
\end{eqnarray}   
\end{linenomath*}       	
Take the derivative with respect to $x_j$, 
\begin{linenomath*}
\begin{equation}
	\partial _j\langle\sigma _{ij}^s\rangle = {\mu _s}(1 - \phi )\partial_j\left\{ \overline {u_{i,j}^s}  
	+ \overline {u_{j,i}^s}  - \frac{2}{3} \delta_{ij}\nabla{\cdot}\overline{\mbox{\boldmath{$u$}}^s}\right\} 
	+ K_s(1 - \phi )\partial_j\nabla {\cdot}\overline{\mbox{\boldmath{$u$}}^s}
	+ \left( K_s- \frac{2}{3}\mu_s \right)\partial_j\left(\phi_{0}-\phi\right).
	\label{3.42}
\end{equation} 
\end{linenomath*}

Now consider the boundary integral term in (\ref{3.23}),
\begin{linenomath*}
\begin{equation}
	J=-\frac{1}{V}\int_{A^{fs}} \sigma_{ij}^s{n_j}dA.
	\label{3.43}
\end{equation} 
\end{linenomath*}
Apply the boundary condition $\sigma _{ij}^s{n_j}=- {p^f}{n_i}+\sigma _{ij}^f{n_j}$ in (\ref{3.43}), 
\[ J=\frac{1}{V}\int_{A^{fs}}\left(p^f n_i-\sigma_{ij}^f n_j\right) dA = I_1+I_2.\]
By (\ref{3.16})--(\ref{3.17}),
\begin{linenomath*}
\begin{equation}
	J=\mu_f(a+ b)\left(\overline{v^f_i}-\overline{v^s_i}\right)
	-\overline{p^f}\,\partial_i\phi.
	\label{3.44}
\end{equation}
\end{linenomath*}

Substituting the relations (\ref{3.27}), (\ref{3.42}) and (\ref{3.44}) in (\ref{3.23}),
and noting that the gradient of porosity is approximately zero, we obtain
\begin{linenomath*}
\begin{eqnarray}
	&&\rho^s \partial_t^2\overline{u_i^s} - \mu_s\partial_j\left[\overline{u_{i,j}^s} + \overline {u_{j,i}^s}
	- \frac{2}{3}{\delta _{ij}}\overline {u_{l,l}^s}\right] -K_s\partial_i\overline {u_{l,l}^s}
	- \frac{{{\mu _f}{\phi ^2}}}{{K(1-\phi )}}(\overline {v_i^f} - \overline {v_i^s})  \nonumber \\ 
	&&= \rho^s\partial_t^2\overline{u_i^s} - {\mu _s}\left[\partial_j\partial_j\overline{u_i^s} 
	+ \frac{1}{3}\partial_i\partial_j\overline{u_j^s}\right] -K_s\partial_i\partial_l\overline{u_l^s} 
	- \frac{{{\mu _f}}}{K}\frac{{{\phi ^2}}}{{1 - \phi }}(\overline {v_i^f} - \overline {v_i^s})  \nonumber \\ 
	&&= 0.
	\label{3.45}
\end{eqnarray}
\end{linenomath*} 	 	 	
The equation is rewritten in a vector form as 
\begin{linenomath*}
\begin{equation}
	\rho^s\partial_t^2\overline{\mbox{\boldmath{$u$}}^s} - {\mu _s}\nabla^2\overline{\mbox{\boldmath{$u$}}^s}
	- (K_s+\frac{\mu_s}{3})\nabla (\nabla{\cdot}\overline{\mbox{\boldmath{$u$}}^s})
	- \frac{\mu_f}{K}\frac{\phi^2}{1 - \phi}(\overline{\mbox{\boldmath{$v$}}^f} - \overline{\mbox{\boldmath{$v$}}^s}) = 0.
	\label{3.46}
\end{equation} 	
\end{linenomath*} 	 	
So, equations (\ref{3.45}) and (\ref{3.46}) are the macro-scale solid equations.

\subsection{Macro-scale equations of motion in tight gas and oil media}\label{subsec:34}

So far, we have derived the following macro-scale equations of solid and fluid for the 
tight gas and oil media (i.e., equations (\ref{3.20}) and (\ref{3.45})):
\begin{linenomath*}
\begin{eqnarray*}
	\partial_t\left(\rho^f\overline{v_i^f}\right)+\partial_j\left(\overline{p^f}\delta_{ij} 
	- \mu_f\left(\partial_i\overline{v_j^f}  +\partial_j\overline{v_i^f}  
	- \frac{2}{3}\delta_{ij}\partial_l\overline{v_l^f}\right)\right) 
	+ {\mu _f}\frac{\phi}{K}\left(\overline{v_i^f} -\overline{v_i^s}\right) &= 0,
	\label{3.47}\\
	\rho^s\partial_t^2\overline{u_i^2}  -\mu_s\left(\partial_j\partial_j\overline{u_i^s} 
	+ \frac{1}{3}\partial_i\partial_j\overline {u_j^s}\right) 
	- {K_s}{\partial_i}{\partial_l}\overline {u_l^s}-\frac{\mu_f}{K}\frac{\phi^2}{1 - \phi}
	\left( \overline {v_i^f}  - \overline {v_i^s} \right) &= 0,
	\label{3.48}
\end{eqnarray*}
\end{linenomath*} 
or, in vector form (i.e., equations (\ref{3.21}) and (\ref{3.46})),
\begin{linenomath*}
\begin{eqnarray*}
	\rho^f\partial_t^2\overline{\mbox{\boldmath{$u$}}^f} + \nabla\overline{p^f}
	- \mu_f\left(\frac{1}{3}\nabla \nabla{\cdot}\overline{\mbox{\boldmath{$v$}}^f}
	+ {\nabla^2}\overline{\mbox{\boldmath{$v$}}^f} \right)
	+{\mu_f}\frac{\phi}{K}\left(\overline{\mbox{\boldmath{$v$}}^f}-\overline{\mbox{\boldmath{$v$}}^s}\right)& = 0,
	\label{3.49}\\
	\rho^s\partial_t^2\overline{\mbox{\boldmath{$u$}}^s} 
	-\mu_s\nabla^2\overline{\mbox{\boldmath{$u$}}^s}  
	- \left(K_s+\frac{\mu_s}{3}\right)\nabla \left( \nabla{\cdot}\overline{\mbox{\boldmath{$u$}}^s} \right) 
	- \frac{\mu_f}{K}\frac{\phi^2}{1-\phi}\left(\overline{\mbox{\boldmath{$v$}}^f}-\overline{\mbox{\boldmath{$v$}}^s}\right)& = 0.
	\label{3.50}
\end{eqnarray*}
\end{linenomath*} 

To close the system of partial differential equations for the motions in the tight oil and gas
media, we need to establish a relation between $\overline{p^f}$ and $\overline{\mbox{\boldmath{$u$}}^f}$. 
Such a relation was discussed in a number of publications, cf. \citep{DS85, MM1999, Spa2001}. 
Here, we follow \citep{DS85} and pay particular attention to the characteristics of tight oil and gas.
We ignore the heat transfer, and thus the entropy does not change in time. In this situation, 
the equation of continuity of fluid is \citep{Landau1987}
\begin{linenomath*}
\begin{equation}
	\frac{\partial\rho^f}{\partial t}+ \nabla  {\cdot} \left(\rho^f\mbox{\boldmath{$v$}}^f\right) = 0.
	\label{3.51}
\end{equation}
\end{linenomath*} 
Take a volume-averaging on both sides of the equation (\ref{3.51}), 
\begin{linenomath*}
\begin{equation}
	\frac{1}{V}\int_V\left\{\frac{\partial\rho^f}{\partial t}+ \nabla{\cdot} \left(\rho^f\mbox{\boldmath{$v$}}^f\right)
	\right\} dV = 0.
	\label{3.52}
\end{equation}
\end{linenomath*} 
Apply the averaging formulas,
\begin{linenomath*}
\begin{eqnarray}
	\frac{1}{V}\int_V \frac{\partial\rho^f}{\partial t} dV 
	&=& \frac{\partial}{\partial t}\left\langle \rho^f\right\rangle 
	- \frac{1}{V}\int_{A^{fs}}\rho^f\mbox{\boldmath{$v$}}^f{\cdot} \mbox{\boldmath{$n$}}\, dA,
	\label{3.53}\\
	\left\langle \nabla{\cdot}\rho^f\mbox{\boldmath{$v$}}^f \right\rangle  
	&=& \nabla{\cdot}\left\langle\rho^f\mbox{\boldmath{$v$}}^f\right\rangle  
	+ \frac{1}{V}\int_{A^{fs}} \rho^f\mbox{\boldmath{$v$}}^f {\cdot} \mbox{\boldmath{$n$}}\, dA.
	\label{3.54}
\end{eqnarray}	
\end{linenomath*}
Using the relations (\ref{3.53}) and (\ref{3.54}) in (\ref{3.52}), we get
\begin{linenomath*}
\begin{equation}
	\frac{\partial}{\partial t}\left(\phi\,\overline{\rho^f} \right)
	+ \nabla {\cdot}(\phi\,\overline{\rho^f}\overline{\mbox{\boldmath{$v$}}^f}) = 0.
	\label{3.55}
\end{equation}
\end{linenomath*}

When the seismic wave is used in oil and gas exploration, the source is loaded near the surface, and
therefore the seismic wave in the reservoirs is a far field one. The size of the strain in a far field wave
(denoted by $\varepsilon$ in the following) is far less than $10^{-9}$ \citep{WN79}.

Let $\rho^{f}$ and $\phi$ be the density of the fluid and porosity of the equilibrium state before 
the seismic waves arrive. In deriving the macro-scale equations with the averaging formulas, 
we may choose the size of the averaging volume suitably so that $\rho^{f}$ and $\phi$ could be 
viewed as constant in the averaging volume. Let $\rho^{f\prime}(t)$ and $\phi^\prime (t)$ be 
perturbations of the density of the fluid and the porosity induced by the waves, and let 
$\rho_{to}^f$ and $\phi_{to}$ be the density of the fluid and the total porosity after the waves arrive. Then,
\begin{linenomath*}
\begin{eqnarray}
	\rho_{to}^f &=& \rho^{f} + \rho^{f\prime}(t),\quad \rho^{f} \gg \rho^{f\prime}(t),
	\label{3.56}\\
	\phi_{to}  &=& \phi+ \phi^\prime (t), \quad \phi\gg \phi^\prime(t).
	\label{3.57}
\end{eqnarray}
\end{linenomath*}

Now we analyze the perturbation size of the fluid density and the porosity.
Consider a small volume element $V_0$ in the oil and gas media.  The volume strain $Q$ is defined as	
\[ Q = \frac{V-V_0}{V_0} = \frac{\Delta V}{V_0} \le 3\varepsilon, \]
where $V$ is the volume element after the perturbation. The volume strain in far field region
induced by the seismic wave is far less than $10^{-6}$. For the small volume element under consideration,
the porosity is just one of the component (the other one is the matrix).  Therefore, for the 
perturbation of the porosity, $\phi^\prime(t) \le Q \ll 10^{-6}$.

Regarding the perturbation of the fluid density induced by seismic waves, due to the conservation of mass, 
we have
\[ \rho^fV_0^f = \left( \rho^{f} + \rho ^{f\prime} \right)\left(V_0^f + \Delta V^f\right),\]
and then,
\[ \left| \frac{\rho^\prime(t)}{\rho} \right| = \left| - \frac{\Delta {V^f}}{V_0^f} \right| \ll 10^{-6},\]
where $V_0^f$ represents the volume of the fluid.

Substituting equation (\ref{3.56}) and (\ref{3.57}) into equation (\ref{3.55}), we have
\begin{linenomath*}
\begin{equation}
	\frac{\partial}{\partial t}\left( \rho^{f} \phi+ \phi\overline{\rho^{f\prime}} +\rho^f{\phi^\prime}  
	+ \phi^\prime\overline{\rho^{f\prime}}\right) + \nabla{\cdot}\left[(\rho^f \phi
	+\phi \overline{\rho^{f\prime}} + \rho^f\phi^\prime  + \phi^\prime \overline{\rho^{f\prime}})
	\overline{\mbox{\boldmath{$v$}}^f}\right] = 0.
	\label{3.58}
\end{equation}
\end{linenomath*}	
Dividing the equation by $\rho^f{\phi}$, we have
\[ \frac{\partial }{\partial t}\left(\frac{\overline{\rho^{f\prime}}}{\rho^{f}} +\frac{\phi^\prime}{\phi} 
+ \frac{\phi^\prime\overline{\rho^{f\prime}}}{\rho^{f}\phi} \right) 
+ \nabla{\cdot} \left[\left(1 + \frac{\overline{\rho^{f\prime}}}{\rho^{f}} +
\frac{\phi^\prime}{\phi}+\frac{\phi^\prime\overline{\rho^{f\prime}}}{\rho^{f}\phi}\right)
\overline{\mbox{\boldmath{$v$}}^f}\right] = 0.\]
Omitting high order small quantities, and noting that $\phi^\prime$ and $\overline{\rho^{f\prime}}$ 
are independent of the space variable for the selected size of the averaging volume suitably, we get
\begin{linenomath*}
\begin{equation}
	\frac{\partial}{\partial t}\left( \frac{\overline{\rho^{f\prime}}}{\rho^{f}}
	+ \frac{\phi^\prime}{\phi} \right) + \nabla\cdot\overline{\mbox{\boldmath{$v$}}^f}
	= \frac{\partial}{\partial t}\left(\frac{\overline{\rho^{f\prime}}}{\rho^{f}}
	+ \frac{\phi^\prime}{\phi}+\nabla \cdot\overline{\mbox{\boldmath{$u$}}^f} \right) = 0.
	\label{3.59}
\end{equation}
\end{linenomath*}
For the wave motion, it implies
\begin{linenomath*}
\begin{equation}
	\frac{\phi^\prime}{\phi}= -\nabla \cdot\overline{\mbox{\boldmath{$u$}}^f}
	- \frac{\overline{\rho^{f\prime}}}{\rho^{f}}.
	\label{3.60}
\end{equation}
\end{linenomath*}
In the following, the unperturbed pressure is taken to be zero. We have \citep{Ber1980,DS85}
\begin{linenomath*}
\begin{equation}
	\overline{\rho^{f\prime}} = \frac{\partial\rho^f}{\partial p^f} \,\overline{p^f} 
	= \frac{1}{c^2}\,\overline{p^f} ,
	\label{3.61}
\end{equation}
\end{linenomath*}
where $c$ is the wave propagation velocity in acoustic media. Use (\ref{3.61}) in (\ref{3.60}), 
\begin{linenomath*}
\begin{equation}
	\frac{\phi^\prime}{\phi} =- \nabla \cdot\overline{\mbox{\boldmath{$u$}}^f}- \frac{1}{c^2\rho^{f}}\,\overline{p^f}.
	\label{3.62}
\end{equation}
\end{linenomath*}
Let
\begin{linenomath*}
\begin{equation}
	\eta (t) = \frac{\phi^\prime}{\phi}.
\end{equation}
\end{linenomath*}
Then,
\begin{linenomath*}
\begin{equation}
	\overline{p^f}=-\rho^{f} c^2 \left(\nabla\cdot\overline{\mbox{\boldmath{$u$}}^f}+ \eta (t)\right).
	\label{3.63}
\end{equation}
\end{linenomath*}
If the solid matrix were completely rigid, then $\eta (t)=0$. The system of equations for the motion in tight oil and gas media consists of (\ref{3.20}), (\ref{3.21}) 
and (\ref{3.63}) or of (\ref{3.21}), (\ref{3.46}) and (\ref{3.63}).

\subsection{The equations of motion corresponding to multi-subdomains}\label{subsec:35}

The macro-scale equations in Subsection~\ref{subsec:34} were derived based on the four basic 
physical assumptions.  If these basic assumptions are not satisfied on the entire domain, then
we may split the domain into several sub-domains so that on each sub-domain, the four basic
physical assumptions are valid.  In this way, we obtain the motion equations for each sub-domain.

For convenience, we consider the case of two sub-domains.

\subsubsection{The boundary conditions on the interface between two sub-domains}
Introduce the following symbols:

$\overline{u_i^{f,1}}$: the $i$th component of fluid average displacement in sub-domain 1

$\overline{u_i^{f,2}}$: the $i$th component of fluid average displacement in sub-domain 2   

$q_i^1$: the $i$th component of fluid  filtering velocity in sub-domain 1

$q_i^2$: the $i$th component of fluid  filtering velocity in sub-domain 2

$\overline{\sigma _{ij}^{f,1}}$: fluid macro-scale stress tensor in sub-domain 1

$\overline{\sigma _{ij}^{f,2}}$: fluid macro-scale stress tensor in sub-domain 2

$\overline{\sigma _{ij}^{s,1}}$: solid macro-scale stress tensor in sub-domain 1 

$\overline{\sigma _{ij}^{s,2}}$: solid macro-scale stress tensor in sub-domain 2 

$q_n$: normal component of the filtering velocity

$\overline{p^1}$: average fluid pressure in sub-domain 1 

$\overline{p^2}$: average fluid pressure in sub-domain 2 

\begin{figure}
	\centerline{\includegraphics[width=0.5\textwidth]{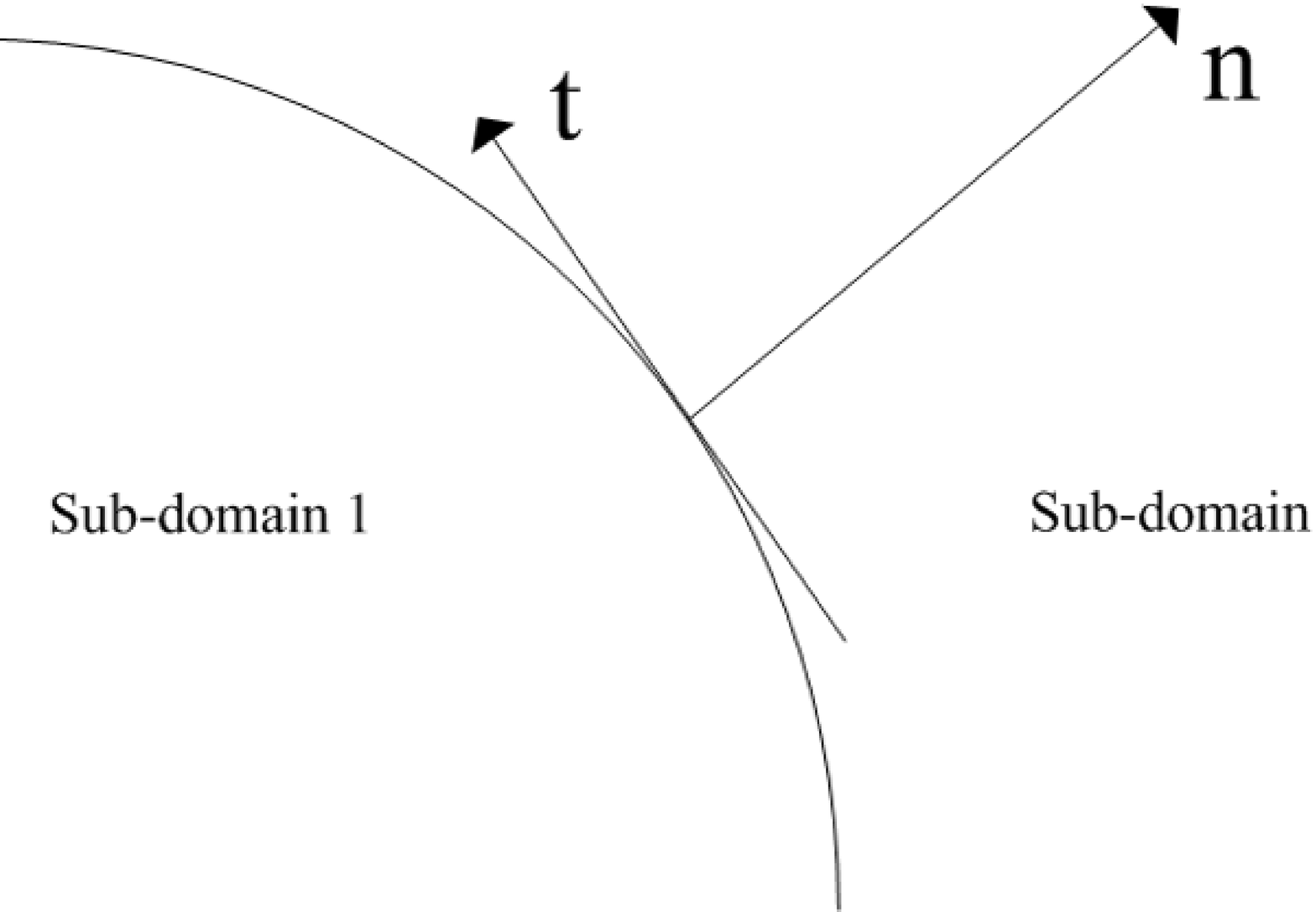}}
	\caption{Boundary between two sub-domains}
	\label{fig1}
\end{figure}

The joining conditions across the boundary of two sub-domains are as follows \citep{BCZ87}.

(1) The macro-scale displacements of solid and fluid are continuous,
\[ u_i^{f,1}=u_i^{f,2}, \quad u_i^{s,1}=u_i^{s,2}.\]

(2) Components of filtering speed are continuous
\[ q_i^1{n_i} = q_i^2{n_i}.\]

(3) Macro-scale stresses on the interface are continuous
\[ \overline{\sigma _{ij}^{f,1}}n_j = \overline{\sigma _{ij}^{f,2}}{n_j},\quad
\overline{\sigma _{ij}^{s,1}}{n_j} = \overline{\sigma _{ij}^{s,2}}{n_j}.\]

(4) The relation between normal component of filtering speed and pressure of two sides on the interface is
\[ q_n=- {\kappa _s}\left(\overline{p^2} -\overline{p^1} \right),\]
where $\kappa_s$ represents fluid permeability of per unit length on the interface.

\subsubsection{Choice of characteristic length for volume averaging in sub-domain}

\begin{figure}
	\centerline{\includegraphics[width=0.5\textwidth]{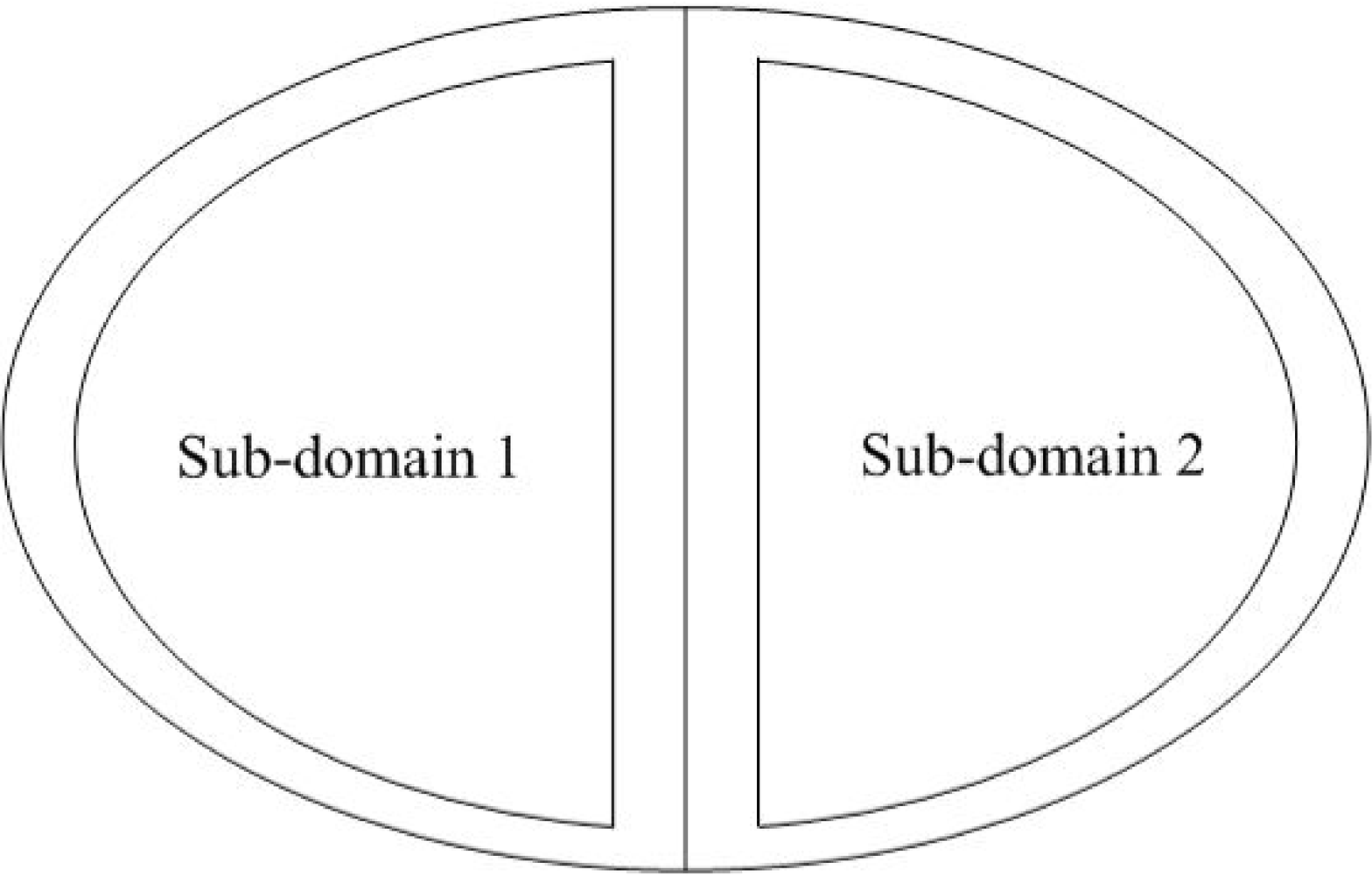}}
	\caption{A domain consists of two sub-domains}
	\label{fig2}
\end{figure}

Figure \ref{fig2} shows the domain splitting into two sub-domains. When applying volume average formulas
on the macro-scale equations of motion of the fluid and solid, the diameter of the balls in volume average
(the characteristic length) can be chosen through the spatial distribution of the geometric construction 
and physical property in order to satisfy the four assumptions of tight oil and gas. For different 
sub-domains, the characteristic length is to be chosen suitably. Based on this, we may view the derivation
method presented in this paper as a volume average method with variable characteristic scale.

\subsubsection{Macro-scale equations of motion of fluid and solid in the entire domain}

The macro-scale equations of motion of solid and fluid in every sub-domain and the jointing conditions 
on the boundaries between sub-domains, form the system of equations of motion in the entire domain.

\section{Governing Equations of Compressional/Shear Waves and Diffusive-Viscous Equation} \label{sec:4}
We begin with the macro-scale equations of motion of tight oil and gas, and derive the governing equations 
of compressional and shear waves. Based on the basic assumption of the tight oil and gas media, 
we simplify the equation of compressional waves and obtain a simplified equation in the form
of the conventional diffusive-viscous wave equation.

We rewrite the macro-scale equations of motion as follows:
\begin{linenomath*}
\begin{eqnarray}
\rho^f\partial_t^2\overline{\mbox{\boldmath{$u$}}^f}- \rho^{f}{c^2}\nabla \left( 
\nabla {\cdot}\overline{\mbox{\boldmath{$u$}}^f}-\eta (t)\right) 
- \mu_f\left(\frac{1}{3}\nabla \nabla {\cdot}\overline{\mbox{\boldmath{$v$}}^f}+ \nabla^2\overline{\mbox{\boldmath{$v$}}^f}\right) 
+ \mu_f\frac{\phi}{K}\left(\overline{\mbox{\boldmath{$v$}}^f}-\overline{\mbox{\boldmath{$v$}}^s}\right)&= 0,
\label{4.1}\\
\rho^s\partial_t^2 \overline{\mbox{\boldmath{$u$}}^s}-\mu_s\nabla^2\overline{\mbox{\boldmath{$u$}}^s}
- \left(K_s+\frac{\mu_s}{3}\right)\nabla\left(\nabla{\cdot}\overline{\mbox{\boldmath{$u$}}^s}\right) 
-\frac{\mu_f}{K}\frac{\phi^2}{1-\phi}\left(\overline{\mbox{\boldmath{$v$}}^f}-\overline{\mbox{\boldmath{$v$}}^s}\right)
&= 0.
\label{4.2}
\end{eqnarray}
\end{linenomath*}
Equation (\ref{4.1}) is derived from (\ref{3.20}) and (\ref{3.63}).

\subsection{Governing equation of compressional wave in tight oil and gas}

Denote
\begin{linenomath*}
\begin{eqnarray}
\varepsilon_f &\equiv \nabla  \cdot\overline{\mbox{\boldmath{$u$}}^f},
\label{4.3}\\
{\varepsilon _s} &\equiv \nabla  \cdot \overline{\mbox{\boldmath{$u$}}^s}.
\label{4.4}
\end{eqnarray}
\end{linenomath*}

Apply the divergence operator on the equation (\ref{4.1}) to get 
\begin{linenomath*}
\begin{equation}
\rho^f\partial_t^2{\varepsilon_f} =\rho^{f}{c^2}\nabla^2{\varepsilon_f} 
+\frac{4}{3}{\mu _f}{\nabla ^2}\frac{\partial\varepsilon_f}{\partial t}
- \mu_f\frac{\phi}{K}\frac{\partial}{\partial t}\left(\varepsilon_f-\varepsilon_s\right).
\label{4.5}
\end{equation}
\end{linenomath*}
Note that ${\eta (t)}$ does not depend on the space variable, and so its divergence is zero.

Apply the divergence operator on the equation (\ref{4.2}) to get
\begin{linenomath*}
\begin{equation}
\rho^s\partial_t^2\varepsilon_s= \left(K_s + \frac{4}{3}\mu_s\right)\nabla^2\varepsilon_s
+ \frac{\mu_f}{K}\frac{\phi^2}{1-\phi}\frac{\partial}{\partial t}\left(\varepsilon_f-\varepsilon_s\right).
\label{4.6}
\end{equation}
\end{linenomath*}

Equations (\ref{4.5}) and (\ref{4.6}) form a governing system of the compressional wave in tight
oil and gas media.

\subsection{Governing equation of compressional wave and diffusive-viscous equations}

According to \citep{SUD02},
\begin{linenomath*}
\begin{equation}
\varepsilon_s\equiv \nabla {\cdot} \overline{\mbox{\boldmath{$u$}}^s}=- \frac{\rho_{to}^s-\rho _0^s}{\rho_0^s},
\label{4.7}
\end{equation}
\end{linenomath*}
where $\rho_0^s$ is the solid density before the wave arrives, $\rho_{to}^s$ is the solid density 
after the wave arrives, $\rho_{to}^s=\rho _0^s+\rho^{s\prime}$. For the tight oil and gas media, 
the bulk modulus of the matrix is two or three orders bigger in magnitude than that of the gas, 
and thus the bulk strain of the matrix may be regarded as zero.  Then (cf. Appendix C),
\begin{linenomath*}
\begin{equation}
\frac{\partial \varepsilon_s}{\partial t} = 0,
\label{4.8}
\end{equation}
\end{linenomath*}
and (\ref{4.5}) is simplified to
\begin{linenomath*}
\begin{equation}
\rho^f\partial_t^2{\varepsilon_f} =\rho^{f}{c^2}\nabla^2{\varepsilon_f} 
+\frac{4}{3}{\mu _f}{\nabla ^2}\frac{\partial\varepsilon_f}{\partial t}
- \mu_f\frac{\phi}{K}\frac{\partial\varepsilon_f}{\partial t}.
\label{4.9}
\end{equation}
\end{linenomath*}
Equation (\ref{4.9}) is the simplified form of governing equation in tight oil and gas media. 
It is of the same form as diffusive-viscous wave equations. We may derive from (\ref{4.9}) that
\begin{linenomath*}
\begin{equation}
\frac{\partial^2\varepsilon_f}{\partial t^2}+\frac{\mu_f}{K\rho^f}\phi\frac{\partial\varepsilon_f}{\partial t}
- \frac{4}{3}\frac{\mu_f}{\rho^f}\nabla^2\frac{\partial\varepsilon_f}{\partial t}
- c^2\nabla^2\varepsilon_f = 0.
\label{4.11}
\end{equation}
\end{linenomath*}

Recall that the diffusive-viscous wave equation is \citep{GK00, KGDS04, ZGL14, ZGZ14}
\begin{linenomath*}
\begin{equation}
\frac{\partial^2\varepsilon_f}{\partial t^2} +\gamma \frac{\partial\varepsilon_f}{\partial t}
- \eta \nabla^2\frac{\partial\varepsilon_f}{\partial t}- \nu^2\nabla^2\varepsilon_f= 0.
\label{4.10}
\end{equation}
\end{linenomath*}

Comparing equations (\ref{4.10}) and (\ref{4.11}), we obtain 
\begin{linenomath*}
\begin{equation}
\gamma =\frac{\mu_f}{K\rho^f}\phi, \quad\eta=\frac{4}{3}\frac{\mu_f}{\rho^f},\quad \nu=c.
\label{4.12}
\end{equation}
\end{linenomath*}
These relations display physical meanings of the parameters of the diffusive-viscous equation.

\subsection{Governing equation of shear wave in tight oil and gas media}

Denote
\begin{linenomath*}
\begin{eqnarray}
\mbox{\boldmath{$\omega$}}_f&=\nabla \times\overline{\mbox{\boldmath{$u$}}^f},
\label{4.13}\\
\mbox{\boldmath{$\omega$}}_s&= \nabla  \times \overline{\mbox{\boldmath{$u$}}^s}.
\label{4.14}
\end{eqnarray}
\end{linenomath*}
Apply the curl operator to the equation (\ref{4.1}), and make use of the equations (\ref{4.10}) and (\ref{4.11}),
\begin{linenomath*}
\begin{equation}
\rho^f\partial_t^2\mbox{\boldmath{$\omega$}}_f-\mu_f\nabla^2\frac{\partial\mbox{\boldmath{$\omega$}}_f}{\partial t}
+ \mu_f\frac{\phi}{K}\frac{\partial}{\partial t}(\mbox{\boldmath{$\omega$}}_f-\mbox{\boldmath{$\omega$}}_s) = 0.
\label{4.15}
\end{equation}
\end{linenomath*}
Similarly, from the equation (\ref{4.2}),  we find
\begin{linenomath*}
\begin{equation}
\rho^s\partial_t^2\mbox{\boldmath{$\omega$}}_s-\mu_s\nabla^2\mbox{\boldmath{$\omega$}}_s
-\frac{\mu_f}{K}\frac{\phi^2}{1-\phi}\frac{\partial}{\partial t}(\mbox{\boldmath{$\omega$}}_f-\mbox{\boldmath{$\omega$}}_s) = 0.
\label{4.16}
\end{equation}
\end{linenomath*}

Equations (\ref{4.15}) and (\ref{4.16}) form a system governing the shear wave in tight oil and gas media.

\section{Comparison with the Conventional Equations for Porous Media}\label{sec:5}
The macro-scale equations of motion based on averaging of the micro-scale equations of solid and fluid
match that in Biot's theory \citep{DS85, PGM92, Sah01, Spa2001}. In the following, we make a
comparison with the results reported in \citep{DS85}.

\subsection{Comparison of motion equations}

The macro-scale equations of motion of solid and fluid found in \citep{DS85} are
\begin{linenomath*}
\begin{eqnarray}
&& \rho^f\frac{\partial}{\partial t}\overline{\mbox{\boldmath{$v$}}^f}
+ \nabla {\overline{p^f}} -\mu_f\left[\frac{1}{3}\nabla \left(\nabla{\cdot}\overline{\mbox{\boldmath{$v$}}^f}\right)
+ \nabla^2\overline{\mbox{\boldmath{$v$}}^f}\right]+\mu_f\frac{\phi}{K}\left(\overline{\mbox{\boldmath{$v$}}^f}-\overline{\mbox{\boldmath{$v$}}^s}\right) = 0,
\label{5.1}\\
&& \rho^s\frac{\partial^2}{\partial t^2}\overline{\mbox{\boldmath{$u$}}^s} - \mu_s[\nabla^2\overline{\mbox{\boldmath{$u$}}^s} 
+ \frac{1}{3}\nabla (\nabla {\cdot}\overline{\mbox{\boldmath{$u$}}^s})] 
- {K_s}\nabla (\nabla {\cdot}\overline{\mbox{\boldmath{$u$}}^s}) \nonumber \\
&&\quad\quad\quad\quad\quad\quad\quad\quad\quad\quad\quad\,\,+ \frac{K_s+ \overline{p^f}}{1 - \phi}\nabla \phi  
- \frac{\mu_f}{K}\frac{\phi^2}{1 - \phi}\left(\overline{\mbox{\boldmath{$v$}}^f}-\overline{\mbox{\boldmath{$v$}}^s}\right)=0,
\label{5.2}\\
&&\overline{p^f}= - \frac{Q}{\phi}\nabla {\cdot}\overline{\mbox{\boldmath{$u$}}^s} 
- \frac{R}{\phi }\nabla{\cdot}\overline{\mbox{\boldmath{$u$}}^f},
\label{5.3}
\end{eqnarray}
\end{linenomath*}
where $Q$ and $R$ are parameters in Biot's theory.

The equations presented in this paper are 
\begin{linenomath*}
\begin{eqnarray}
\rho^f\partial_t^2\overline{\mbox{\boldmath{$u$}}^f}-\rho^{f} c^2\nabla \nabla{\cdot}
\overline{\mbox{\boldmath{$u$}}^f} -\mu_f\left(\frac{1}{3}\nabla \nabla{\cdot}
\overline{\mbox{\boldmath{$v$}}^f}+ \nabla^2\overline{\mbox{\boldmath{$v$}}^f}\right) 
+ \mu_f\frac{\phi}{K}(\overline{\mbox{\boldmath{$v$}}^f}-\overline{\mbox{\boldmath{$v$}}^s}) = 0,
\label{5.4}\\
\rho^s\partial_t^2\overline{\mbox{\boldmath{$u$}}^s} -\mu_s\nabla^2\overline{\mbox{\boldmath{$u$}}^s}- 
\left(K_s + \frac{1}{3}{\mu_s}\right)\nabla\left(\nabla{\cdot} \overline{\mbox{\boldmath{$u$}}^s}\right) 
- \frac{\mu_f}{K}\frac{\phi^2}{1-\phi}(\overline{\mbox{\boldmath{$v$}}^f} - \overline{\mbox{\boldmath{$v$}}^s}) = 0.
\label{5.5}
\end{eqnarray}
\end{linenomath*}

As the averaging volume is selected suitably so that the gradient of the porosity may be viewed zero
in the averaging volume, we obtain a closed form for the averaged pressure of the fluid and 
simplified equations of motion.

\subsection{Comparison of governing wave equations}

The governing equations of the shear wave found in \citep{DS85} are
\begin{linenomath*}
\begin{eqnarray}
&\rho^f\partial_t^2\mbox{\boldmath{$\omega$}}_f- \mu_f\nabla^2\partial_t\mbox{\boldmath{$\omega$}}_f
+ \mu_f\frac{\phi}{K}\partial_t(\mbox{\boldmath{$\omega$}}_f- \mbox{\boldmath{$\omega$}}_s) = 0,
\label{5.6}\\
&\rho^s\partial_t^2\mbox{\boldmath{$\omega$}}_s- \mu_s\nabla^2\mbox{\boldmath{$\omega$}}_s
-\frac{\mu_f}{K}\frac{\phi^2}{1-\phi}\partial_t(\mbox{\boldmath{$\omega$}}_f-\mbox{\boldmath{$\omega$}}_s) = 0.
\label{5.7}
\end{eqnarray}
\end{linenomath*}
These equations are identical to our governing equations of shear wave of tight oil and gas media.

The governing equations of the compressional wave found in \citep{DS85} are
\begin{linenomath*}
\begin{eqnarray}
\rho^f\partial_t^2\varepsilon_f &=& \frac{Q}{\phi}{\nabla^2}{\varepsilon_s} 
+ \frac{R}{\phi}{\nabla ^2}{\varepsilon_f} + \frac{4}{3}\mu_f\nabla^2\partial_t\varepsilon_f 
- \mu_f\frac{\phi}{K}\partial_t\left(\varepsilon_f-\varepsilon_s \right),
\label{5.8}\\
\rho^s\partial_t^2{\varepsilon_s} &=& \frac{\phi}{1-\phi}\left(K_s-\frac{K_sR}{\phi\rho^f c^2} \right)
\nabla^2\varepsilon_f+ \left(K_s+ \frac{4}{3}\mu_s-\frac{K_sQ}{\left(1-\phi\right)\rho^f c^2} \right)
\nabla^2\varepsilon_s \nonumber \\
&&+ \frac{\mu_f}{K}\frac{\phi}{1-\phi}\partial_t\left(\varepsilon_f-\varepsilon_s\right).
\label{5.9}
\end{eqnarray}
\end{linenomath*}

The corresponding equations derived in this paper are
\begin{linenomath*}
\begin{eqnarray}
 \rho^f\partial_t^2\varepsilon_f &=& \rho^{f} c^2\nabla^2\varepsilon_f+\frac{4}{3}\mu_f\nabla^2
\partial_t\varepsilon_f-\mu_f\frac{\phi}{K}\partial_t\left(\varepsilon_f-\varepsilon_s\right),
\label{5.10}\\
 \rho^s\partial_t^2\varepsilon_s &=& \left(K_s+ \frac{4}{3}\mu_s\right)\nabla^2\varepsilon_s
+ \frac{\mu_f}{K}\frac{\phi^2}{1-\phi}\partial_t\left(\varepsilon_f-\varepsilon_s\right).
\label{5.11}
\end{eqnarray}
\end{linenomath*}
We observe that our equations for the compressional wave are simpler than those in \citep{DS85}.

\section{Conclusions and Discussions}\label{sec:6}
\subsection{Conclusions}
We present the equations of motion for solid and fluid in tight oil/gas reservoir based on the four basic
assumptions. We derive the macro-scale motion equations through the averaging technique applied on 
the micro-scale motion equations for both the fluid and the solid.  The derivation is relatively 
rigorous.  The equations provide a reliable basis for study of problems in tight oil/gas exploration.

The derived motion equations are more concise than that in Biot's theory, and are better suited for
parameter inversion of porous media.

Our equations lead to the diffusive-viscous equation in the study of compressional waves.
Traditionally, the diffusive-viscous equation is proposed based on experiments, and the 
physical meanings of the parameters of the equation are not clear.  Our derivation provides
relations between the parameters and physical quantities of the media such as the density, 
the porosity, and the effective permeability. This makes it possible to invert the
physical quantities of the media through solving inverse problems.
\subsection{Discussion}
1. The motion equations in tight oil/gas sandstone media are presented under the four assumptions. The 
equations are obtained by employing the volume averaging to the motion equations and the boundary 
conditions on the fluid-solid interface governing each phase at micro-scale. The proposed equations are two 
coupled vector equations in which the parameters have definite physical interpretation. 

2. The validity of the four assumptions in tight oil/gas sandstone media is demonstrated by measured 
petrophysical parameters, the rock physics measurements and the results of the previous studies. 

3. The compressional wave equation is obtained by introducing the scalar potential function . It has the
same form as the conventional diffusive-viscous wave equation, but the parameters in the compressional 
wave equation are
determined by the petrophysical parameters and they have precise physical meaning. Moreover, the derived 
compressional wave equation is easily applicable to seismic inversion due to its simplified form.
 
4. The shear wave equation can be derived by introducing vector potential function in the proposed 
motion equations, which is of important significance in multi-wave inversion. 

\appendix
\section{Appendix A.  Derivation of a formula for $a+b$}

In order to use the observed physical parameter to determine the value of $a+b$, we rewrite
the equation (\ref{3.6}) as
\begin{linenomath*}
\begin{equation}
\partial_t\langle\rho^f v_i^f\rangle + \partial_i\langle p^f\rangle 
+ \partial_j\left[\langle \rho^f v_i^fv_j^f\rangle -\langle\sigma_{ij}^f\rangle \right] 
+ \frac{1}{V}\int_{A^{fs}}\left(p^f\delta_{ij} - \sigma_{ij}^f\right) n_j dA = 0.
\label{B-1}
\end{equation}
\end{linenomath*}

In equation (\ref{B-1}), the first term of left side is the inertial force; the second term is the force generated by the gradient of average pressure; the third term is quadratic term of velocity, which, namely, refers to the force generated by the gradient of space derivation of velocity; the fourth term is the viscous force, which is generated from the coupling between the fluids and solid frame. The experimental data suggest that the velocity of fluid in the tight oil/gas sandstone may be relatively high, while the fluid flux of passing through porous area is small, namely, the averaging velocity of fluid is low. It is because that viscous fluid moves slowly, the time derivative of average velocity (the first term of equation (\ref{B-1})), and second-order partial derivative of fluid’s velocity to space variable, and the gradient of average value of second term of fluid’s velocity are also high-order infinite quantity comparing with both the second and fourth term. As a result, they can be neglected \citep{PGM92}.

Following \citep{PGM92}, we may ignore higher order terms and approximate (\ref{B-1}) by
\begin{linenomath*}
\begin{equation}
{\partial_i}\left(\phi\,\overline{p^f} \right) + \frac{1}{V}\int_{A^{fs}} p^f n_i dA 
- \frac{1}{V}\int_{A^{fs}} \sigma_{ij}^f n_i dA  =0.
\label{B-3}
\end{equation}
\end{linenomath*}
Substituting the equation (\ref{3.16}) and (\ref{3.17}) to (\ref{B-3}) and assuming the gradient 
of the porosity is negligible, we obtain
\begin{linenomath*}
\begin{equation}
\partial_i\left(\phi\,\overline{p^f} \right)+\mu_f a\left(\overline{v_i^f} - \overline{v_i^s}\right)
+\mu_f b\left(\overline{v_i^f} - \overline{v_i^s} \right)=0,
\label{B-4}
\end{equation}
\end{linenomath*}
i.e.,
\begin{linenomath*}
\begin{equation}
\phi \nabla\overline{p^f} = -(a + b)\mu_f\left(\overline{\mbox{\boldmath{$v$}}^f}-\overline{\mbox{\boldmath{$v$}}^s} \right).
\label{B-5}
\end{equation}
\end{linenomath*}
Let $\mbox{\boldmath{$q$}} $ be the filtering speed, defined as \citep{DS83}
\begin{linenomath*}
\begin{equation}
\mbox{\boldmath{$q$}} = \frac{1}{V}\int_V \left(\mbox{\boldmath{$v$}}^f- \mbox{\boldmath{$v$}}^s \right) dV 
= \phi \left(\overline{\mbox{\boldmath{$v$}}^f} -\overline{\mbox{\boldmath{$v$}}^s} \right).
\label{B-6}
\end{equation}
\end{linenomath*}
Substituting equation (\ref{B-6}) into (\ref{B-5}),
\begin{linenomath*}
\begin{equation}
\mbox{\boldmath{$q$}} =  - \frac{\phi^2}{(a + b)}\frac{1}{\mu_f}\nabla\overline{p^f}.
\label{B-7}
\end{equation}
\end{linenomath*}
Based on the equation (\ref{B-7}), the equivalent permeability is \citep{DS83}
\begin{linenomath*}
\begin{equation}
K = \frac{\phi^2}{a + b}.
\label{B-8}
\end{equation}
\end{linenomath*}
Then we get
\begin{linenomath*}
\begin{equation}
a + b = \frac{\phi^2}{K}.
\label{B-9}
\end{equation}
\end{linenomath*}

\section{Appendix B. Rationality of the four assumptions}

%\subsection*{The space variation of porosity of the tight oil/gas sandstone}

The statistic data analysis and micro-phase sandstone volume suggest that porosity of 
Sulige tight oil/gas sandstone is primarily controlled by the micro-scale sedimentary phase \citep{Rey99}.
The lateral resolution can be divided into internal 
sandstone block of river channel, margin sandstone block of river channel, internal floodplain of 
river channel and different micro-scale phase types.

We take the data from the Erdos basin as an example.

Internal sandstone block of river channel: within the width of 2-4 km of micro-scale phase of river 
channel, the porosity generally ranges from 6\% to 8\%, while the lateral variation of porosity is less than 
1\% scaling from 10 m to 1000 m.

Margin sandstone block of river channel: the width of margin of sandstone block is generally less than
1 km with the porosity ranging from 3\% to 6\%, and its lateral variation of porosity is within 1\%--2\%
at the lateral scaling from 100 m to 1000 m. The lateral variation is slightly high.

Internal floodplain of river channel: the width of internal floodplain is 2--4 km, and the porosity 
is less than 3\%. The lateral porosity variation is less than 1\% at the scaling from 10 m to 4000 m.

The porosity variation of different micro-scale phase: the maximum lateral variation of porosity 
is observed between the micro-scale phase of river channel and that of different river channel, 
the distance among them is generally less than 1 km, while the porosity can be varied from 8\% to 2\%, 
namely, the lateral porosity variation is 6\% per thousand meters. 

When calculating the average volume, the characteristic length is more than $10^{-6}$ m and 
less than dozens of meters. Therefore, choosing the characteristic length of average volume suitably, 
the gradient of the porosity may be regarded as zero.

Table \ref{tab:ModelASymbol1} provides numerical ranges of the P wave velocity, the S wave velocity 
and the density in tight gas sandstone for the case of dry saturation
with ultrasonic measurement at an effective pressure of 40\,Mpa \citep{Jiz1991}.

\begin{table}
\caption{Ranges of P wave velocity, S wave velocity and density in tight gas sandstone}
\centering
\label{tab:ModelASymbol1}
\begin{tabular}{ccc}
\hline
  & Minimum & Maximum\\
\hline
Vp (km/s) & 3.81 & 5.57 \\
Vs (km/s) & 2.59 & 3.50 \\
Density(g/cc${}^2$) & 2.26 & 2.67 \\
\hline
\end{tabular}
\end{table}

Table \ref{tab:ModelASymbol2} provides a classification of pore sizes \citep{Sch2011}.  The pore size
in tight oil/gas reservoir is typically less than $50\,\mu$\,m.

\begin{table}[th]
\centering
\caption{Pore size classification}
\label{tab:ModelASymbol2}
\begin{tabular}{cccc}
\hline
Type of pore  & Diameter $d$ & Type of Pore & Diameter $d$\\
\hline
Rough pore & $d > 2\,{\rm mm}$ & Macropore & $50\,{\rm nm} > d > 2\,{\rm nm}$ \\
Macrocapillary & $2\,{\rm mm} > d > 50\,\mu{\rm m} $ & Mesopore & $2\,\mu {\rm m} > d > 0.8\,{\rm nm}$ \\
Capillary & $50\,\mu {\rm m} > d > 2\,\mu {\rm m}$ & Micropore  & $0.8\,{\rm nm} > d$\\
Microcapillary & $2 \,\mu {\rm m}> d > 5\,{\rm nm}$ & & \\
\hline
\end{tabular}
\end{table}

\section{Appendix C.  Demonstration of $\partial_t\varepsilon_s=0$}\label{App-c}

When the pressure induced by the seismic wave propagates into the region, the density of the solid will be 
changed. Recall equation (\ref{4.8}):
\begin{linenomath*}
\begin{equation}
\varepsilon_s(\mbox{\boldmath{$x$}},t) =  - \frac{\rho^s(\mbox{\boldmath{$x$}},t) -\rho_0^s(\mbox{\boldmath{$x$}})}{\rho_0^s( \mbox{\boldmath{$x$}})}.
\label{D-1}
\end{equation}
\end{linenomath*}
Differentiate with respect to $t$,
\begin{linenomath*}
\begin{equation}
\frac{\partial\varepsilon_s}{\partial t} = - \frac{1}{\rho^s_0}\frac{\partial\rho^s}{\partial t}.
\label{D-3}
\end{equation}
\end{linenomath*}

The bulk modulus of the matrix in tight oil and gas (rock matrix) is related to the porosity 
\citep{PW2005}. For sandstone, the bulk modulus of different porosity are shown as

\begin{table}[th]
\centering
\caption{The bulk modulus of different porosity}
\label{tab:ModelASymbol3}
\begin{tabular}{ll}
\hline
Porosity & Bulk modulus $K_s$ (GPa)\\
\hline
0.06 & 36 \\
0.18 & 38.6 \\
0.19 & 36 \\
\hline
\end{tabular}
\end{table}

Because most gases are extremely compressible under reservoir conditions,
in many cases the bulk modulus (incompressibility) of a hydrocarbon gas can be set as 0.01-0.2 GPa 
in seismic modeling \citep{Sch2011}. Let ${K_g}$ represent the bulk modulus of gas, 
based on the data mentioned above.  We have
\begin{linenomath*}
\begin{equation}
\frac{K_s}{K_g} \sim {10^2} - {10^3}.
\end{equation}
\end{linenomath*}
The bulk modulus of the matrix in tight oil and gas is 2 to 3 orders bigger in magnitude than that of the gas, 
the bulk strain of solid matrix resulting from the pressure induced by seismic waves may be zero, 
that is, the change in the density is negligible.   Hence,
\[ \frac{\partial\rho^s(\mbox{\boldmath{$x$}},t)}{\partial t} =0, \]
and by (\ref{D-3}), 
\[ \frac{\partial\varepsilon_s}{\partial t} =0.\]

\acknowledgments
The research reported in the paper was supported by the Major Program of National Natural Science Foundation 
of China under grant No.\ 41390454, the National Science and Technology Major Projects under grant 
No.\ 2016ZX05024-001-007 and No.\ 2017ZX050609, the Major Research Plan of the National Natural Science 
Foundation of China under grant No.\ 91330204, the National Natural Science Foundation of China under 
grant No.\ 41504091, the projects of National Engineering Laboratory for Offshore Oil Exploration. The data provided in Table~\ref{tab:ModelASymbol1} is available by contacting the corresponding author (Jinghuai Gao) at jhgao@mail.xjtu.edu.cn. The data provided in Table~\ref{tab:ModelASymbol2} and Table~\ref{tab:ModelASymbol3} is available from \citet{Sch2011} and \citet{PW2005}, respectively.

%% ------------------------------------------------------------------------ %%
%% Citations

% Please use ONLY \citet and \citep for reference citations.
% DO NOT use other cite commands (e.g., \cite, \citeyear, \nocite, \citealp, etc.).

%% Example \citet and \citep:
%  ...as shown by \citet{Boug10}, \citet{Buiz07}, \citet{Fra10},
%  \citet{Ghel00}, and \citet{Leit74}.

%  ...as shown by \citep{Boug10}, \citep{Buiz07}, \citep{Fra10},
%  \citep{Ghel00, Leit74}.

%  ...has been shown \citep [e.g.,][]{Boug10,Buiz07,Fra10}.

%%  REFERENCE LIST AND TEXT CITATIONS
%
% Either type in your references using
%
% \begin{thebibliography}{}
% \bibitem[{\textit{Kobayashi et~al.}}(2003)]{R2013} Kobayashi, T.,
% Tran, A.~H., Nishijo, H., Ono, T., and Matsumoto, G.  (2003).
% Contribution of hippocampal place cell activity to learning and
% formation of goal-directed navigation in rats. \textit{Neuroscience}
% 117, 1025--1035.
%
% \bibitem{}
% Text
% \end{thebibliography}
%
%%%%%%%%%%%%%%%%%%%%%%%%%%%%%%%%%%%%%%%%%%%%%%%
% Or, to use BibTeX:
%
% Follow these steps
%
% 1. Type in \bibliography{<name of your .bib file>}
%    Run LaTeX on your LaTeX file.
%
% 2. Run BiBTeX on your LaTeX file.
%
% 3. Open the new .bbl file containing the reference list and
%   copy all the contents into your LaTeX file here.
%
% 4. Run LaTeX on your new file which will produce the citations.
%
% AGU does not want a .bib or a .bbl file. Please copy in the contents of your .bbl file here.

\bibliography{my.bib}

%% After you run BibTeX, Copy in the contents of the .bbl file here:

%%%%%%%%%%%%%%%%%%%%%%%%%%%%%%%%%%%%%%%%%%%%%%%%%%%%%%%%%%%%%%%%%%%%%
% Track Changes:
% To add words, \added{<word added>}
% To delete words, \deleted{<word deleted>}
% To replace words, \replaced{<word to be replaced>}{<replacement word>}
% To explain why change was made: \explain{<explanation>} This will put
% a comment into the right margin.

%%%%%%%%%%%%%%%%%%%%%%%%%%%%%%%%%%%%%%%%%%%%%%%%%%%%%%%%%%%%%%%%%%%%%
% At the end of the document, use \listofchanges, which will list the
% changes and the page and line number where the change was made.

% When final version, \listofchanges will not produce anything,
% \added{<word or words>} word will be printed, \deleted{<word or words} will take away the word,
% \replaced{<delete this word>}{<replace with this word>} will print only the replacement word.
%  In the final version, \explain will not print anything.
%%%%%%%%%%%%%%%%%%%%%%%%%%%%%%%%%%%%%%%%%%%%%%%%%%%%%%%%%%%%%%%%%%%%%

%%%
\listofchanges
%%%

\end{document}